\documentclass[aps,prl,twocolumn,superscriptaddress,groupedaddress]{revtex4}
\usepackage{subfig}
\usepackage{graphicx}  
\usepackage{dcolumn}   
\usepackage{bm}        
\usepackage{amssymb}   
\usepackage{slashed}
\usepackage{graphicx}				
\usepackage{amsmath}
\usepackage{mathtools}
\usepackage{tikz,pgf}
\usepackage{comment}
\usepackage{asymptote}
\usetikzlibrary{arrows,backgrounds}
\usetikzlibrary{fit,scopes,calc,matrix,positioning,decorations.pathmorphing}
\usepackage[all]{xy}
\usepackage{yfonts}
\newcommand{\bra}[1]{\ensuremath{\left\langle#1\right|}}
\newcommand{\ket}[1]{\ensuremath{\left|#1\right\rangle}}
\newcommand{\Bracket}[1]{\ensuremath{\left\langle#1\right\rangle}}

\begin{document}
\title{Holographic quantum information and (de)confinement}
\author{Andrei T. Patrascu}
\address{ELI-NP, Horia Hulubei National Institute for R\&D in Physics and Nuclear Engineering, 30 Reactorului St, Bucharest-Magurele, 077125, Romania\\
email: andrei.patrascu.11@ucl.ac.uk}
\begin{abstract}
Analysing the phenomenon of deconfinement from a holographic point of view, it appears that the brane configuration in the bulk, corresponding to the confinement phase imposes a restriction on the strength of the holographic quantum error correction procedure. This restriction is partially removed when the transition to a deconfined phase occurs. The brane configurations corresponding to the bulk instantons are analysed and it is shown that they cannot reach the region of the bulk that would ensure maximal error protection in the confinement phase, while keeping non-zero instanton size. In the deconfinement phase on the other side, we have a configuration that allows a higher degree of quantum error protection, still preventing the maximal protection allowed by the holographic principle. This suggests that the strength of quantum error correction codes in QCD systems is fundamentally limited both in confinement and deconfinement phases. 
\end{abstract}
\maketitle
\section{Introduction and review of useful concepts}
Following the holographic revolution and the introduction of the AdS/CFT duality, condensed matter physics changed dramatically up to the point where its foundations have been re-written using the language of quantum gravity, and in certain limits, that of general relativity [1]. This duality has allowed us to explore regimes of quantum many body systems that are completely inaccessible by other more conventional means. The holographic principle became particularly pertinent in the physics of non-Fermi liquid states of matter in finite density systems of strongly interacting fermions [2]. There it suggested that a new principle is at work, one that is connected to the physics of compressible quantum matter: the nature of the state of matter is governed, as it seems, by a macroscopic quantum entanglement involving all of its constituents. Large scale entanglement seemed to be particularly expressive in determining what type of quantum states we detect [3]. However, it remained unclear how holography would link to quantum information. The AdS/CFT duality can be derived from string theory in certain limits and this suggested that indeed, maybe general relativity is part of the quantum theory of strings. In a sense this led L. Susskind to claim that general relativity is a description of quantum mechanics. Quantum field theory, as a theoretical construction of its own, is as far removed from general relativity as one can think. While it was clear how to resolve certain quantum field theories in curved background, the actual calculations in the cases when the gravitational field was sufficiently strong that second order effects would become relevant became soon intractable. The old problem of non-renormalisability of gravity comes into mind here. However, Maldacena's discovery connected quantum field theory and general relativity in a beautiful way. He showed that in certain limits, these two theories can be two aspects of the same theory. General relativity and quantum field theory seem to be dual to each other in a deep and absolute way [4]. After the discovery of the duality, Gubster, Klebanov, Polyakov, and Witten (GKPW) discovered a set of mathematical rules showing how one can relate results in the gravitational side of the duality with results in the conformal field theory side of it [5]. But at the fundamental level, one has to remember that AdS/CFT is actually relating stringy quantum gravity to certain quantum field theories. The string theory side of the duality remains notoriously hard to understand as string theory is extremely difficult to work with. However it seems very likely that many aspects of physics at various scales are controlled by the physics of quantum strings. If we consider certain limits, stringy quantum gravity becomes basically a solution of general relativity and that can directly be linked to field theory. These limits are by themselves interesting. They specifically require that the field theory would have matrix valued fields of rank $N$ and that one considers the regime of large $N$ and strong coupling. The standard example would be a very strongly coupled $SU(N)$ Yang-Mills theory with a very large number of colours. Within this limit one may look for solutions using general relativity and re-trace them in quantum field theory by means of our holographic dictionary. We would certainly like to consider a method to free ourselves from the large N limit but when this is being tried, the difficulties of quantum gravity become manifest and the results are hard to control. 
\par However, there are various situations where this large $N$ limit is not that much of a problem. We have two related concepts, the one called UV-independence in string theory which is equivalent with "strong emergence" as used by condensed matter physicists. Both underline the independence of macroscopic phenomena from the underlying microscopic degrees of freedom. The origin of this observation can be related to the construction of statistical physics by Boltzmann at the end of the nineteenth century. This led to the description of the basic phases of thermal matter and the generalisation to the zero temperature quantum matter in the form of Landau style order parameter theories describing superfluids or superconductors. The culmination of this reasoning was the description of Fermi liquids. However the next step to be taken was that of the Wilsonian renormalisation group. Holography is capable of expressing the resulting strongly emergent theories of matter by means of the language of general relativity. This leads to the possibility of predicting the states of matter that are ruled by macroscopic quantum entanglement. The renormalisation group refers to the property of a field theory that, by tracing out the short distance degrees of freedom one induces a flow in the parameters of the theory that describes how the theory changes as one lowers the scale. This scaling direction turns out to be related to the extra geometrical dimension of the gravity dual. The scaling flow of the field theory becomes fully encoded in the geometrical properties of the higher dimensional spacetime which is governed by Einstein's field equations. Counting the degrees of freedom in a gravitational theory is equivalent to counting the degrees of freedom of a quantum field theory in a dimension fewer. In the AdS/CFT theory, the first part is represented by the well known Anti-de-Sitter space which represents the general relativity side of the duality. It is a solution of Einstein's equations with negative cosmological constant and it is the Lorentzian higher-dimensional generalisation of a hyperboloid. Light-like propagation will reach the edge of this spacetime in a finite time. Hence the gravity theory will have to be supplemented with a boundary and the information therein. This boundary is where the field theory lives and the RG flow is associated to the extra radial dimension moving towards the centre of the AdS space. The second part is the conformal field theory. Conformal field theories describe the universal behaviour at second order phase transitions. Here however they do appear naturally as zero-temperature relativistic field theories. They are quantum critical theories describing the near zero temperature quantum phase transitions controlled by another external parameter. 
Basically the holographic principle tells us that the quantum theory of gravity can be written in one fewer spatial dimensions. However we know that low energy physics has a simple description in 3+1 dimensions. If these processes should have an equivalent description in a lower dimensional theory we may ask how come that a local quantum field theory in 3+1 dimensions is usually described by a larger dimensional Hilbert space than the 2+1 dimensional one. One may think of a special sector of the quantum field theory which is mapped into a lower dimensional theory. But which one? And how is the mapping working? This is a special question to be asked in quantum information theory where mapping one Hilbert space into another is a fundamental problem about safely encoding information and storing it in a way that is safe against errors [6], [7]. For this to work we may start with a logical Hilbert space $\mathcal{H}_{logical}$ and a physical Hilbert space $\mathcal{H}_{phys}$. The error correcting code is given by an encoding map $E:\mathcal{H}_{logical}\rightarrow \mathcal{H}_{phys}$ and a decoding map $D:\mathcal{H}_{phys}\rightarrow \mathcal{H}_{logical}$. Clearly the dimension of $\mathcal{H}_{phys}$ will be larger than the dimension of $\mathcal{H}_{logical}$ as the logical information to be encoded will use more physical degrees of freedom to safely encode it. The image of the encoding map will define a subspace $\mathcal{H}_{code}$ which is known as the error correcting code. An error correcting code protects agains an error $A$ if for $\rho\in\mathcal{H}_{logical}$ we have that $(D\circ A\circ E)(\rho)=\rho$. A class of error correcting codes is known as the stabiliser codes. They are being defined by a stabiliser group which is a subgroup of the Pauli group on $n$ qubits. The Pauli group $G_{n}$ is formed from all the n-fold tensor products of Pauli operators. A stabiliser group is given by a set of generators drawn from the Pauli group. A stabiliser code is defined as the joint +1 eigenspace of the generators of the stabiliser group. Consider the physical Hilbert space made up of three qutrits: $\mathcal{H}_{phys}=\mathcal{H}_{1}\otimes \mathcal{H}_{2}\otimes \mathcal{H}_{3}$ with the stabiliser group generated by the elements $S_{1}=X\cdot X\cdot X$ and $S_{2}=Z\cdot Z\cdot Z$ where $X$ and $Z$ are the generalised Pauli operators acting on qutrits. The positive eigenstates will be defined by $S_{i}\ket{\psi}=\ket{\psi}$ for all $i$. The states satisfying this property are
\begin{equation}
\begin{array}{c}
\ket{0_{L}}=\frac{1}{\sqrt{3}}{\ket{000}+\ket{111}+\ket{222}}\\
\ket{1_{L}}=\frac{1}{\sqrt{3}}{\ket{012}+\ket{201}+\ket{120}}\\
\ket{2_{L}}=\frac{1}{\sqrt{3}}{\ket{021}+\ket{102}+\ket{210}}\\
\end{array}
\end{equation}
The subspace defined by the three states $\ket{0_{L}}$,$\ket{1_{L}}$,$\ket{2_{L}}$ generates an error correcting code capable of correcting against the erasure of any one of the qutrits. That means there is an unitary operator $U_{12}$ such that 
\begin{equation}
U_{12}\otimes I \ket{i}_{123}=\ket{i}_{1}\otimes \ket{\chi}_{23}
\end{equation}
where $\ket{\chi}=\frac{1}{\sqrt{3}}(\ket{00}+\ket{11}+\ket{22})$. Therefore one Hilbert space can be encoded into another larger one with some redundancy. This means that some loss of parts of the larger Hilbert space can still allow the recovery of the encoded state without breaking the no-cloning theorem. The AdS/CFT duality states that 
\begin{equation}
Z_{CFT}[J]=Z_{AdS}[\Phi\rightarrow J]
\end{equation}
where the CFT involves some external sources $J$ and the AdS partition function contains fields that have boundary conditions defined by those sources. The quantum error correction code can naturally be transferred into the AdS/CFT context. The Ryu Takaianagi theorem gives us the area of a bulk region $\gamma_{A}$ using the data of the boundary interval $A$. If we focus on subintervals of $A$ we can calculate the area of an infinite family of regions that sweep the entire region enclosed by $\gamma_{A}$. Thus the region enclosed by the minimal surface $\gamma_{A}$ labeled $D(A)$ is the region that can be reconstructed from $A$. We also have to consider the causal bulk domain of dependence of $D(A)$ which we denote $W(A)$ and call the entanglement wedge of $A$. Considering an operator $O_{W(A)}$ at the point $x$ inside the entanglement wedge of $A$, we may use it to probe the bulk region and by the reconstruction theorem there should be a dual operator $O_{A}$ living in the boundary theory with the property that 
\begin{equation}
Tr(\rho_{W(A)}O_{W(A)})=Tr(\rho_{A}O_{A})
\end{equation}
However a single bulk point lies in the entanglement wedge of many different boundary regions the bulk to boundary encoding in AdS/CFT has the structure of an error correcting code. It can also be seen that the deeper the operator lies in the bulk the more it is protected against accidental erasures. 
One can understand this with a simplified $2+1$ dimensional AdS space. Therefore let us define this metric as
\begin{equation}
ds^{2}=l_{AdS}^{2}(-cosh^{2}(\rho)dt^{2}+d\rho^{2}+sinh^{2}(\rho)d\theta^{2})
\end{equation}
with $\rho\in\mathbb{R}^{+}$, $t\in\mathbb{R}^{+}$, $\theta\in[0,2\pi]$, and $l_{AdS}$ being the AdS radius. If we want to look at a constant time slice then it will have the metric $ds^{2}=(d\rho^{2}+sinh^{2}(\rho)d\theta^{2})$. The $AdS_{2+1}$ space can be seen as a cylinder. A convenient coordinate system for the AdS space is the Poincare system. It does not cover the entire space but only a restricted region called the Poincare patch. Its metric is 
\begin{equation}
ds^{2}=\frac{l_{AdS}^{2}}{z^{2}}(-dt^{2}+dz^{2}+dx_{\mu}dx^{\mu})
\end{equation}
where $0\leq z < \infty$, and $t,x\in\mathbb{R}$.The boundary of the AdS space can be seen as the surface of the cylinder. Using global coordinates this will appear at $\rho\rightarrow \infty$ while in Poincare coordinates it appears at $z=0$. In this example the conformal field theory is $1+1$ dimensional. In this scenario the surfaces in the bulk we will be interested in will be spacelike geodesics anchored at two boundary points. These geodesics form semicircles. The length of such a geodesic is 
\begin{equation}
A=\frac{l_{AdS}}{2G_{N}}ln(\frac{L}{\epsilon})
\end{equation}
where L is the size of the boundary interval on which the geodesic is anchored and $z=\epsilon$ defines the cutoff surface as we approach the boundary. The entanglement entropy of a single interval in a $1+1$ dimensional CFT  is given by 
\begin{equation}
S(A)=\frac{c}{3}log(\frac{L}{\epsilon})
\end{equation}
with $c$ the central charge determined by the type of CFT we work with and $\epsilon$ is a cutoff eliminating the modes of wavelength shorter than itself. The length of the boundary anchored geodesics in AdS space and the entanglement entropy of the same boundary interval correspond if 
\begin{equation}
c=\frac{3 l_{AdS}}{2 G_{N}}
\end{equation}
and if the UV cutoff in the CFT is identified with the long distance $z$ cutoff in AdS. The Ryu Takayanagi formula the is conjectured to be 
\begin{equation}
S(A)=\min_{\gamma_{A}}\frac{L(\gamma_{A})}{4G}
\end{equation}
where $\{\gamma_{A}\}$ are all the spacelike curves anchored in region $A$. When additional fields aside of the metric are present in the gravitational bulk the Ryu-Takayanagi formula is modified to take them into account as 
\begin{equation}
S(A)=\min_{\gamma_{A}}\frac{L(\gamma_{A})}{4G}+S(\rho_{W(A)})
\end{equation}
where $S(\rho_{W(A)})$ is the entropy of the fields in the region enclosed by $\gamma_{A}$. However the RT formula depends on the existence of a preferred time slicing of the AdS spacetime. This may not be the same in generalisations to other spacetimes. The minimal surfaces $\gamma_{A}$ should retain their meaning and hence they must lie in a well chosen time slice. If that were not so, i.e. if they could be chosen in any Cauchy slice of AdS with the region $A$ on its boundary one could always take the minimal length as close to zero as one wishes by approaching the slice to being lightlike. For a spacetime with a timelike Killing vector (a static spacetime) and boundary conditions which are at a constant time $t=t_{0}$ the bulk Cauchy slice defined by $t=t_{0}$ contains the appropriate minimal curve. In other situations however the Ryu Takayanagi formula is replaced by the Hubeny-Ranganmani-Takayanagi formula
\begin{equation}
S(A)=extremal_{\gamma_{A}}\frac{L(\gamma^{A})}{4G}
\end{equation}
The procedure of minimisation is therefore replaced with an extremisation. Assuming the null energy condition this formula is equivalent with the so called maximin prescription leading to 
\begin{equation}
S(A)=\max_{\Sigma}(\min_{\gamma_{A}}\frac{L(\gamma^{A})}{4G})
\end{equation}
meaning that for each bulk Cauchy slice $\Sigma$ we search all possible curves $\gamma_{A}$, take out the shortest, and define it to be $\gamma_{\Sigma}$. The length of the longest of the $\gamma_{\Sigma}$ gives the entanglement entropy of $A$. It is therefore suggested by the maximin description of the entanglement entropy that the AdS/CFT duality is an encoding of one Hilbert space into another. Using the equality of partition functions we see how the encoding of the boundary Hilbert space into the bulk Hilbert space occurs. Once we specify the boundary state and solve the bulk equations of motion to determine the bulk fields we can reverse the mapping and take the limits of the bulk fields towards the boundary to recover boundary data. 
If however there is no timelike killing vector then it has been shown that certain tensor networks can represent the features of the geometry of arbitrary Cauchy slices in any asymptotically AdS spacetime [31], [32], [33]. The toy model that could be used is representing this spacetime in a discrete manner being based on tensor networks in which the length on the network is defined by the mutual information. Random tensor networks can further represent the situation in which there is no isometry in the time direction. Indeed the RT formula holds without restrictions on a graph geometry and one is capable of proving the RT formula even when there are no bulk legs available. A network constructed on any graph will display the RT formula and in particular there are no conditions for isometry in the case when no bulk legs are present. To understand this better it is necessary to look at the formulation of tensor networks in holographic contexts. 
The tensor network itself defines the pattern of index contractions of a set of objects that together form a quantum state. Each vertex of the network corresponds to a quantum state and each line corresponds to a ket or bra index. The directionality of the line indicates kets (inward arrows) and bra (outward arrows). A maximally entangled state therefore would correspond to a vertex with two inward oriented arrows (or outwards oriented arrows for the dual representation). Often one writes the quantum states without including their basis vectors in a tensor network hence for a maximally entangled state we may write $\ket{\Psi^{+}}=\delta_{ab}$ leaving the basis choice implicit. Ket indices are lowered and bra indices are raised and hence $\bra{\Psi^{+}}=\delta^{ab}$. In general if we have a quantum state $\ket{\phi}=\sum_{a,b}T_{ab}\ket{a}\ket{b}$ then we can assume it is fully specified by the tensor $T_{ab}$. The fundamental operation of a tensor network is the composition of two quantum states for example $T_{abcd}$ and $S_{efgh}$ as defined by the use of maximally entangled bra states $T_{abcd}\circ S_{efgh}\rightarrow T_{abcd}\delta^{ce}\delta^{df}S_{efgh}$. This is of course not unique as different pairs of indices could be contracted. The contraction pattern of two or more quantum states is described by means of a graph. States are associated with vertices and each line is associated to an index. Contraction means placing maximally entangled pairs on the edges and connecting inward and outward going lines. Operators and observables acting on quantum states are represented as vertices having both inward and outward lines attached hence will be written as tensors with both upper and lower indices. Indices can be raised and lowered by contracting with maximally entangled pairs. Therefore an operator can be mapped to a state $M_{a}^{b}\rightarrow M_{ab}=M_{a}^{c}\delta_{cb}$ and states to operators. The density matrix corresponding to the state $\ket{\phi}=\sum_{a,b}T_{ab}\ket{a}\ket{b}$ is 
\begin{equation}
\rho_{AB}=\sum_{ab}T_{ab}(T^{*})^{cd}\ket{a}\bra{c}_{A}\otimes\ket{b}\bra{d}_{B}
\end{equation}
Partial trace is represented by contraction of the corresponding inward or outward indices. It has been noticed early on by Van Raamsdonk and others that given a thermofield double state 
\begin{equation}
\ket{\Psi}=\sum_{i}e^{-\beta E_{i}/2}\ket{E_{i}}_{A}\otimes \ket{E_{i}}_{B}
\end{equation}
with $A$ and $B$ the Hilbert spaces of the two CFTs this would correspond to a wormhole geometry in the bulk. The entanglement between the two subsystems $A$ and $B$ being responsible for for the bulk wormhole connection between the different spacetimes. One may use the parameter $\beta$ to decrease the entanglement entropy between $A$ and $B$ and observe that the area of the wormhole neck will decrease pinching off the wormhole as the entanglement vanishes. This shows that entanglement builds geometry in the bulk. An isometry group of a manifold is the group of bijective maps that can preserve distance as defined by the respective metric. The infinitesimal generators are the Killing fields and the Lie derivative of the metric along the Killing fields vanishes. For a tensor network we have a cut $\gamma$ that divides the network into two regions say $C$ and $\bar{C}$. On this cut we can define a state $\ket{\Psi}_{B\bar{B}}$. The regions $C$ and $\bar{C}$ have bulk legs associated to the corresponding indices. We can project on those indices and obtain the bulk state $\ket{\Phi}_{W(A)W(\bar{A})}$. A subregion isometry exists if given a cut of minimal length $\gamma_{A}$ anchored on a boundary region $A$, the tensor network defines the maps $C:H_{W(A)B}\rightarrow A$ and $\bar{C}:H_{W(\bar{A})\bar{B}}\rightarrow H_{\bar{A}}$ both being isometries meaning they preserve the distance between the two regions. A network with subregion isometry leads to both the RT formula and to the quantum error correction property. Given such a network we can construct a state 
\begin{equation}
\ket{\Psi}=C\otimes \bar{C}(\ket{\Psi}_{B\bar{B}}\otimes\ket{\Phi}_{W(A)W(\bar{A})})
\end{equation}
The reduced density matrix of a subregion $A$ will be
\begin{equation}
\rho_{A}=Tr_{\bar{A}}(C\otimes \bar{C}(\ket{\Psi}\bra{\Psi}_{B\bar{B}}\otimes\ket{\Phi}\bra{\Phi}_W(A)W(\bar{A}))C^{\dagger}\otimes \bar{C}^{\dagger})
\end{equation}
by the cyclic property of the trace and using $\bar{C}^{\dagger}C=\mathbb{I}$ one derives
\begin{equation}
\rho_{A}=C[Tr_{\bar{B}}(\ket{\Psi}\bra{\Psi}_{B\bar{B}})\otimes Tr_{W(\bar{A})}(\ket{\Phi}\bra{\Phi}_{W(A)W(\bar{A})})]C^{\dagger}
\end{equation}
The von Neumann entropy is not changed under conjugation by an isometry hence
\begin{equation}
\begin{array}{c}
S(\rho_{A})=S(Tr_{\bar{B}}(\ket{\Psi}\bra{\Psi}_{B\bar{B}}))+S(\rho_{W(A)})\\
=|\gamma_{A}|log(D)+S(\rho_{W(A)})\\
\end{array}
\end{equation}
here $|\gamma_{A}|$ is the number of legs cut by $\gamma$ and $D$ is the dimension of a single leg. $|\gamma_{A}| log(D)$ is the length of the cut and hence
\begin{equation}
S(\rho_{A})=\min_{\gamma_{A}}L(\gamma_{A})+S(\rho_{W(A)})
\end{equation}
which is the RT formula including the bulk entropy term. 
The subregion isometry also leads to the error correction property. This property requires that there exists a boundary operator $O_{A}$ on the region $A$ for every bulk operator $O_{W(A)}$ living in the entanglement wedge such that 
\begin{equation}
Tr(\rho_{A}O_{A})=tr(\rho_{W(A)}O_{W(A)})
\end{equation}
In a tensor network these operators can be constructed by defining
\begin{equation}
O_{A}=C(O_{W(A)}\otimes\mathbb{I}_{B})C^{\dagger}
\end{equation}
Then
\begin{widetext}
\begin{equation}
Tr(\rho_{A}O_{A})=Tr(\rho_{A}C(O_{W(A)}\otimes\mathbb{I}_{B})C^{\dagger})=Tr(C^{\dagger}\rho_{A}C(O_{W(A)}\otimes \mathbb{I}_{B}))
\end{equation}
\end{widetext}
and $C^{\dagger}\rho_{A}C=\rho_{W(A)}\otimes\rho_{B}$
leading precisely to the error correction
\begin{widetext}
\begin{equation}
Tr(\rho_{A}O_{A})=Tr((\rho_{W(A)}\otimes\rho_{B})(O_{W(A)}\otimes\mathbb{I}_{B}))=tr(\rho_{W(A)}O_{W(A)})
\end{equation}
\end{widetext}
where use has been made of the fact that both $C$ and $\bar{C}$ are isometries. Various types of networks featuring these properties have been constructed. Such networks are in general toy models displaying features similar to AdS/CFT. The boundary legs associated to the corresponding indices are the CFT degrees of freedom while the tensor networks graph geometry is represented as the geometry of a bulk Cauchy slice. It is known that the basic formulation of the RT formula must be extended when one deals with dynamic spacetimes to the HRT formula or the maximin formula. To incorporate this in a tensor network the definition of length must be modified as compared to the one used for static spacetimes. Dynamical spacetimes can be used as a benchmark to infer the connection between bulk and boundary in the case in which no direct isometry exist between the two regions. Even if the spacetime is static, the tools associated to this development help in linking with a highly non-trivial bulk as the one encoding the phase transition between confinement and deconfinement. In particular, a modification of the error correction capabilities of the bulk can be linked to the existence of a phase transition. Therefore I suspect the possibility of connecting various phases of gauge theories not only with the simple existence of large scale entanglement or other intrinsically material quantities, but also with various measures originating in quantum computing and quantum information, as is the capacity of implementing quantum error correction, discussed in this article. The definition of length used customary in tensor networks, namely 
\begin{equation}
L_{G}(\gamma)=log(dim(\gamma))
\end{equation}
will not be sufficient anymore. The new definition of length will include the notion of a quantum state residing on a cut. From the maximin formula 
\begin{equation}
S(A)=\max_{\Sigma}(\min_{\gamma_{A}}L(\gamma_{A}))
\end{equation}
on each spacelike slice of the boundary $\Sigma$ with the chosen boundary and the subset $A$ of the slice we calculate the length of the minimal surface homologous to $A$ and from the set of all those lengths we choose the largest one. This will generate the entropy $S(A)$. The maximisation step provides us with a bound on the minimal length by the entropy 
\begin{equation}
\min_{\gamma_{A}}L(\gamma_{A})\leq S(A)
\end{equation}
This restricts the possible definitions of the length $L(\gamma_{A})$. The rank bound on the entanglement entropy would ask that 
\begin{equation}
S(A)\leq \min_{\gamma_{A}}(log(dim(\gamma_{A})))=\min_{\gamma_{A}}L_{G}(\gamma)
\end{equation}
This is the exact opposite inequality leaving the sole possibility $S(A)=\min_{\gamma_{A}}L_{G}(\gamma)$ for all the networks in the set we optimise over. Every network in that set must therefore contain the extremal curve anchored in $A$. If we repeat for each of the boundary regions we arrive at the conclusion that every network must contain the extremal curves for each boundary network, which is impossible in a dynamical spacetime where exactly no one slice should contain all of the extremal curves. We therefore need a new definition of the graph length. There indeed exists a definition of $L(\gamma)$ in which $S(A)=L(\gamma_{iso}^{A})$ if an isometric cut exists in the set of networks associated with the boundary state. We first calculate the boundary state on $A$ in terms of the operators defined by an isometric cut $\gamma_{iso}^{A}$.
We obtain
\begin{equation}
\rho_{A}=C Tr_{\bar{B}}(\ket{\Psi}\bra{\Psi}) C^{\dagger}
\end{equation}
But $C$ is an isometry and hence we have $S(\rho_{A})=S(Tr_{\bar{B}}(\ket{\Psi}\bra{\Psi}))$. Any cut has a state associated with it. In particular for an isometric cut we have
\begin{equation}
\ket{\gamma_{iso}}=\ket{\Psi}_{B\bar{B}}
\end{equation}
Therefore defining the length as the entropy of one side of $\ket{\Psi}$ would provide us with $S(A)$ and in a more symmetric form we can define the length by means of mutual information
\begin{equation}
L(\gamma_{iso}^{A})=\frac{1}{2}I_{\ket{\gamma}}(B:\bar{B})
\end{equation}
We allow extra index legs in forming our network because we need them for bulk construction. At this point it is clear how this allows us to deal with time evolutions in the bulk and boundary in a simple way. Let us have a tensor network with the isometric subregion property. We can evolve the bulk state forward in time by applying an operator to a portion of its Hilbert space. We can then project this bulk state to the boundary by projecting it into the tensor network and we can then push through this time evolution operator to the interior legs of the tensor network. The time evolution operator can then be treated as the same type of deformations to the network as the ones considered for dynamical spacetimes and hence local time evolution on the network leads to deformations of the Cauchy slice. This works best for random tensor networks. 
\section{confinement-deconfinement transition and quantum error correction}
The previous section was an important review of some concepts that will clarify the procedures we will perform in what follows and make them well defined. The use of tensor networks as toy models for holography in connection with quantum error correction measures can improve our thinking about the phase structure of a four dimensional SU(N) pure Yang Mills theory at finite temperature and large $N$. The $Z_{N}$ symmetry of this theory along the temporal cycle, which is also the centre of the $SU(N)$ symmetry has as order parameter the temporal Polyakov loop 
\begin{equation}
W_{0}=\frac{1}{N} Tr(P e^{i\int_{0}^{\beta}A_{0}dx^{0}})
\end{equation}
The vanishing of this operator indicates the preservation of the $Z_{N}$ symmetry and hence physical quantities do not depend on the temporal radius $\beta/2\pi$ at $O(N^{2})$. Also, the preservation of the $Z_{N}$ symmetry implies that the confinement phase is dominant, which happens at low temperatures. At high temperatures we have a broken $Z_{N}$ symmetry, a deconfinement phase and $W_{)}\neq 0$. At intermediate temperatures we have therefore to find a phase transition. In the confinement phase the free energy does not depend on temperature and hence the entropy becomes zero at $O(N^{2})$ while in the deconfinement phase the entropy will be $O(N^{2})$. That means that in the confinement phase we have no gluons in the spectrum, which would have $O(N^{2})$ degrees of freedom, but instead we only have gauge singlet states like glueballs leading to $O(1)$ order entropy. Looking at $SU(N)$ pure Yang Mills theory from a holographic point of view, in the bulk we may start from a 10 dimensional Euclidean spacetime with an $S^{1}$ cycle and consider $D4$ branes wrapping this $S^{1}$ cycle. The coordinate around this cycle is $x_{4}$ and the periodicity is $L_{4}$. The effective theory on this brane will be a 5-dimensional supersymmetric Yang-Mills theory on $S^{1}_{L_{4}}$. If we consider fermions on the brane the boundary conditions can be antiperiodic or periodic. The choice of antiperiodic boundary conditions will produce fermion masses according to the Kaluza Klein mechanism proportional to the scale $1/L_{4}$ and that will produce supersymmetry breaking according to the Scherk-Schwarz mechanism inducing masses for the adjoint scalars and for $A_{4}$ proportional to $\lambda_{4}/L_{4}$ up to one loop. Given the dynamical scale $\Lambda_{YM}=\Lambda exp[-1/(b\lambda_{4})]$ and $\lambda_{4}$ defined as the value of the coupling at the cut-off scale $\Lambda$ while $b$ is the first coefficient of the $\beta$ function, then if $\lambda_{4}$ is small and the dynamical scale is much smaller than the mass scales the fermions and adjoint scalars decouple and we reduce the 5-dimensional supersymmetric Yang-Mills theory to a 4-dimensional Yang Mills theory. In the large $N$ limit at low temperatures the gravity dual of the compactified SYM theory consists of a solitonic D4 brane solution wrapping around the $S_{L_{4}}^{1}$. The metric is
\begin{widetext}
\begin{equation}
\begin{array}{c}
ds^{2}=\alpha'[\frac{u^{3/2}}{\sqrt{d^{4}\lambda_{5}}}(dt^{2}+\sum_{i=1}^{3}dx_{i}^{2}+f_{4}(u)dx_{4}^{2})+\frac{\sqrt{d_{4}\lambda_{5}}}{u^{3/2}}(\frac{du^{2}}{f_{4}(u)}+u^{2}d\Omega_{4}^{2})]\\
\\
f_{4}(u)=1-(\frac{u_{0}}{u})^{3}\\
\\
e^{\phi}=\frac{\lambda_{5}}{(2\pi)^{2}N}(\frac{u^{3/2}}{\sqrt{d_{5}\lambda_{5}}})^{1/2}\\
\end{array}
\end{equation}
\end{widetext}
where $\lambda_{5}=\lambda_{4}\cdot{L_{4}}$ is the Yang Mills coupling on the D4 world-volume and $d_{4}$ is found as the particular case $p=4$ in
\begin{equation}
d_{p}=2^{7-2p}\pi^{9-3p)/2}\Gamma(\frac{7-p}{2})
\end{equation}
and in order to avoid conical singularities as our $x_{4}$ cycles can be reduced to zero for $u=u_{0}$ the asymptotic radius must be chosen such that 
\begin{equation}
\frac{L_{4}}{2\pi}=\frac{\sqrt{d_{4}\lambda_{5}}}{3}u_{0}^{-1/2}
\end{equation}
We compactify Euclidean time in the boundary theory on a circle with periodicty $\beta=1/T$ while keeping the temperature way below the Kaluza-Klein scale $T<<\lambda_{4}/L_{4}$. Fixing the periodicity of the fermions in the gauge theory along the time cycle determines the gravitational theory. Along the $(t,x_{4})$ directions we can fix periodic or anti-periodic fermionic boundary conditions. Considering gravity theory compactified on $S_{\beta}^{1}\times S_{L_{4}}^{1}$ and choosing $(P,AP)$ boundary conditions for the fermions we obtain a non-trivial phase boundary clarifying the transition between confinement and deconfinement. It has been promoted in [8] that the phase structure of the 5-dimensional SYM on $S_{\beta}^{1}\times S_{L_{4}}^{1}$ with $(P,AP)$ boundary conditions has a strong coupling region described by type $II$ supergravity and characterised by a Gregory-Laflamme phase transition occurring at a temperature 
\begin{equation}
\beta_{GL}=\frac{C_{3}}{C_{4}}\frac{L_{4}}{\lambda_{4}}\cong 8.54\frac{L_{4}}{\lambda_{4}}
\end{equation}
In the weak coupling region the 4-dimensional Yang Mills is realised at low temperatures $(\beta>>\frac{L_{4}}{\lambda_{4}})$. Although this region is common to the $(P,AP)$ region and to the $(AP,AP)$ region the mirror under $Z_{2}(\beta\leftrightarrow L_{4})$ does not exist since the $(P,AP)$ boundary condition is not $Z_{2}$ symmetric. The resulting localised solitonic D3 phase has the same order parameters as the deconfinement phase and hence these two phases may be smoothly connected. In the proposal of [8] the Gregory-Laflamme transition in the $IIB$ frame has been identified with the confinement deconfinement transition in gauge theory. In the Gregory-Laflamme transition, the distribution of the D3 branes on the dual circle changes from a uniform distribution at low temperatures to a localized one at high temperatures. Since the 5D gauge theory of the D4 branes is a IIA theory, to transition to D3 branes we consider a T-duality along the temporal direction. Considering this, $X_{0}$ being transverse to the D3 brane is mapped to the gauge potential $A_{0}$ on the D4 brane. The D3 brane distribution on the dual circle is related to the eigenvalues of the Polyakov loop operator 
\begin{equation}
\begin{array}{cc}
exp(i\theta_{k})\delta_{kl}=[P exp(i\int_{0}^{\beta}A_{0}dt)]_{kl}, & k,l=1,...,N
\end{array}
\end{equation}
If the D3 branes are uniformly distributed, then we can consider $\theta_{k}=2\pi k/N$ and the temporal Polyakov loop operator becomes
\begin{equation}
W_{0}=\frac{1}{N}Tr P e^{i\int_{0}^{\beta}A_{0}dx^{0}}=\frac{1}{N}\sum_{k=1}^{N}e^{2\pi i k/N}=0
\end{equation}
Hence this distribution corresponds to the confinement phase. If the D3 branes are localised then 
$\theta_{k}$ are also localised and the Polyakov loop becomes non-zero characterising the deconfinement phase. This shows a connection between the D3 distribution and the eigenvalue distribution of the Polyakov loop operator. From the connectivity point of view delocalisation vs. localisation of the D3 branes will be related to an entanglement structure reaching or not reaching deeper into the bulk and hence to a different level of quantum error correction. That this is the case will be shown in what follows.

\section{Quantum error correction and phase transition in QCD}
Quantum chromodynamics (QCD) while the best theory of quarks and gluons in existence, shows us still many mysteries. At a first glance, it is not a conformal theory, and hence, the application of the best known practical realisation of the holographic principle, the AdS/CFT duality [1] seems difficult if not impossible. However, the ability of extracting information about a strongly coupled four-dimensional gauge theory by looking at a gravitational theory in five dimensional spacetime is particularly enticing [2]. This becomes even more important when one notices that the low energy regime of QCD is the strongly coupled one, and at the same time, the closest to our empirical scrutiny. Its probing through a gauge-gravity duality would lead to its mapping into a well understood gravitational problem. The AdS/QCD duality is currently an incomplete correspondence that tries to apply a five-dimensional theory on an anti-de-Sitter gravity background in order to produce new results regarding QCD [3]. The exact gravity dual for QCD however is currently unknown. Thanks to 't Hooft's idea, the large-N QCD can be reformulated as a string theory [4] giving us enough room to analyse various QCD phenomena both from a gauge and from a (quantum) gravitational perspective. At low energy, both gauge theory and string theory provides us with a natural form of confinement. In ref. [5] it has been shown that the stringy behaviour and the theories used in the holographic approach provide us with a low-energy behaviour compatible with QCD. Given a truly conformal field theory, the dual string spacetime would be 
\begin{equation}
ds^{2}=\frac{r^{2}}{R^{2}}\eta_{\mu\nu}dx^{\mu}dx^{\nu}+\frac{R^{2}}{r^{2}}dr^{2}+R^{2}ds_{X}^{2}
\end{equation}
where we have considered as in [5] the product of an $AdS_{5}$ space with a five dimensional transverse space $X$. The AdS space has a radius given by $R$ and the transverse, bulk coordinate (with its zero starting on the boundary) is denoted by $r$. If the boundary gauge theory is non-conformal, as is the case for QCD, the same spacetime still corresponds to the gauge theory in the large $r$ limit, but for small $r$, if there is a mass gap, there is a nonzero lower bound $r_{min}$ in the bulk. If the lightest glueball state defines a scale $\Lambda$ then the cutoff $r_{min}$ will have the form 
\begin{equation}
r_{min}\sim \Lambda R^{2}
\end{equation}
From the perspective of QCD the fifth coordinate corresponds to the energy scale too, but higher energy QCD physics is reflected by the behaviour of fields closer to the AdS boundary, where we can still analyse the theory from a gauge theoretical perspective [6]. To make the theory confining, one has to introduce an IR cutoff in the metric, where spacetime ends, in analogy with the cascading gauge theory of ref. [7]. This way of introducing confinement by placing a cut-off in the bulk space, has profound implications in the quantum error correction interpretation of holography.

In order to understand how quantum error correction emerges in a holographic context one may consider looking at the AdS/CFT duality and use a tensor network structure to cover the bulk as shown in ref. [29]. The tensor used for the construction of the network is one with maximal entanglement along any bipartition, giving rise to an isometry from the bulk Hilbert space to the boundary Hilbert space. Such a construction makes the entire tensor network in the bulk an encoder for a quantum error correction code. The bulk degrees of freedom are identified as the logical degrees of freedom capable of robustly encoding the physical degrees of freedom on the boundary. The bulk therefore contains additional layers all contributing towards the error-safety of the boundary encoded information. 
The additional bulk degrees of freedom represented in this context the respective information required to robustly encode the physical information. The non-uniqueness of the boundary CFT operators corresponding to a single bulk operator is interpreted as showing the fact that the bulk operator is a logical operator that is capable of preserving a code subspace of the Hilbert space of the boundary CFT. Such a code subspace is protected against errors that would erase information from the boundary. The larger the metric distance between the bulk operator and the boundary becomes, the more it is protected against erasures on the boundary. On the opposite side, the closer an operator is to the boundary, the smaller the area vital to its reconstruction on the boundary will be, and the more sensitive it will be to accidental erasures on the boundary.
Therefore the ability of the bulk to implement error corrections is given by the bulk depth of the network. For the Anti-de-Sitter spacetime this would amount to the radius of the bulk space. However, in physical situations we do not expect to continue having an Anti-de-Sitter space everywhere in the bulk. More natural situations would involve various String-Brane structures for which a practical dual is not necessarily known. The case of QCD is however partially (and certainly incompletely) described by a system of D1-D3 branes which are going to be regarded here. While the complications related to such a system are of course great, what matters from the perspective of the error correction is that the D-branes involved will induce certain limitations in the ability to reach an arbitrary inner bulk region with an unmodified tensor network. It therefore appears that the description of a confined QCD phase differs from the description of a deconfined QCD phase and both differ from a normal AdS bulk spacetime by the fact that certain inner regions won't be accessible to the tensor network bulk expansion and hence will limit the ability of performing the full quantum error correction of which the AdS space would be otherwise capable. This is particularly important as for the first time it is possible to link the quantum error correction capability to QCD phases and phase transitions. 

The holographic principle seems to be the common denominator linking both QCD and quantum error correction codes to gravity theories. There are however several possible classes of tensor networks more or less suitable for the geometry one wishes to use. It is well known that the so called HaPPY networks [30] have the property that cuts crossing a minimal number of network edges saturate an entropy bound. This allows HaPPY networks to precisely connect graph geometry and entanglement and to automatically satisfy the Ryu-Takayanagi formula. The essential part of this type of network is the perfect tensor. This tensor defines a unitary transformation from any set of $n$ network edges to their complement. Therefore perfect tensors define isometries from any subset of edges of size $k<n$ to their complement. Clearly the geometry associated to QCD is different, particularly because the D-branes involved in the bulk are pinching off at a coordinate distance in the bulk, say $u=u_{0}$. In order for this to happen, we have a sharper pinch-off of the surfaces and hence the tensor network required to follow this will cut through a smaller number of network edges. 

 Analysing the emergence of confinement from this point of view brings new insights both for the understanding of confinement to deconfinement phase transition, and to the understanding of the power of holographic quantum error correction codes. While this description is certainly illuminating, its applicability is only formal, as the actual bulk space has a more complex D-brane structure. This structure can, in certain cases, make regions of the bulk inaccessible to a tensor network expansion, therefore limiting the power of quantum error correction. It is important to underline that there is a deep connection between the measures provided by quantum computing and quantum information to determine what operations can be performed robustly on a quantum computer and how such operations can be constructed in an error protected way, and various phase transitions in gauge systems like Yang Mills of QCD. This article wishes to be only the first making probably the simplest connection evident. Many more such connections between quantum information properties and phases of matter should become manifest in the future.

Let us consider an euclidean 10-dimensional spacetime with the $x_{4}$ direction compactified on a circle with period $L_{4}$. The confinement phase in holographic QCD can be described in bulk theory by means of a D-brane construction involving instantons [9-12]. Following reference [9] consider a compactified 5D SYM theory on D4-branes having at low temperature a solitonic D4-brane solution. The explicit metric and the dilaton is
\begin{widetext}
\begin{equation}
\begin{array}{c}
ds^{2}=\alpha'[\frac{u^{3/2}}{\sqrt{\lambda_{5}/4\pi}}(dt^{2}+\sum_{i=1}^{3}dx_{i}^{2}+f_{4}(u)dx_{4}^{2})+\frac{\sqrt{\lambda_{5}/4\pi}}{u^{3/2}}(\frac{du^{2}}{f_{4}(u)}+u^{2}d\Omega_{4}^{2})]\\
\\
f_{4}(u)=1-(\frac{u_{0}}{u})^{3}\\
\\
e^{\phi}=\frac{\lambda}{(2\pi)^{2}N}(\frac{u^{3/2}}{\sqrt{\lambda_{5}/4\pi}})^{1/2}\\
\end{array}
\end{equation}
\end{widetext}

 The system will consist of N D4 branes and D0 branes corresponding to instantons winding on the $x_{4}$ circle. The DBI action for a single instanton in the D4-brane geometry is given by [13,14]
\begin{equation}
S_{D0}=\frac{8\pi^{2}N}{\lambda_{YM}}\sqrt{1-\frac{u_{0}^{3}}{u^{3}}}
\end{equation}
Here $u$ is the position of the instantonic D0-brane along the radial bulk coordinate. 
The first theoretical model linking a gravitational theory to low energy QCD (or Yang Mills) in a phenomenologically meaningful way was the so called Sakai-Sugimoto model. The development of holographic QCD models has been a continuous effort over several decades. Among the first attempts was the Kruczenski-Mateos-Myers-Winters (KMMW) model [38] that introduced probe D6 branes in a D4 background compactified on a circle with supersymmetry breaking boundary conditions and explored various aspects of low energy QCD phenomena. However, they did not obtain the massless pions as Nambu-Goldstone bosons associated to the spontaneous breaking of the $U(N_{f})_{L}\times U(N_{f})_{R}$ chiral symmetry. The Sakai-Sugimoto model introduces instead D8 branes in the same D4 background [39]. The brane configuration in this case in the weakly coupling regime is then given by $N_{c}$ D4 branes compactified on a supersymmetry breaking $S^{1}$ and $N_{f}$ $D8-\bar{D8}$ pairs transverse to this $S^{1}$. This system is T-dual with the $D3/(D9-\bar{D9})$ system considered in ref. [40] with the difference that on the cycle $S^{1}$ on which the T-duality is taken the imposed boundary conditions are anti-periodic for the fermions on the $D4$ branes so that supersymmetry is broken and zero mass fields become massive. 
The $U(N_{f})_{L}\times U(N_{f})_{R}$ chiral symmetry in QCD is realised as a gauge symmetry of the $N_{f}$ $D8-\bar{D8}$ pairs. The radial coordinate $U$ in this model, transverse to the $D4$ branes is bounded from below due to the existence of a horizon in the supergravity background for $U\geq U_{KK}$. As the coordinate $U$ gets closer to $U_{KK}$ the radius of the circle $S^{1}$ goes to zero and the $D8-\bar{D8}$ branes merge at a point $U=U_{0}$ and form a single component of the $D8$ branes. This leads to a single $D8$ brane with only one $U(N_{f})$ factor which is interpreted as a holographic dual for the spontaneous chiral symmetry breaking. The Nambu-Goldstone bosons of this spontaneous symmetry breaking are associated to the massless pions and this theory did contain massless pseudo-scalar mesons in the meson spectrum. The low energy effective action of the massless pion was shown to be identical with that of the Skyrme model. The Skyrme term was introduced to stabilise the soliton solution of a non-linear sigma model, and the protected soliton, called Skyrmion represents a baryon. The Skyrmion has been identified in this model with the D4 brane wrapped around the $S^{4}$ cycle, constructed as a soliton in the world-volume gauge theory of the probe $D8$ branes. The wrapped $D4$ brane is the Witten baryon vertex to which one can attach $N_{c}$ fundamental strings and is a colour singlet bound state of the $N_{c}$ fundamental quarks. However, despite its advantages, this model is ultimately incorrect. First, the holographic model based on this brane system is not equivalent with the large $N_{c}$ QCD in the high energy regime. The four dimensional theory is obtained by compactifying D4 branes to a circle of radius $M_{KK}^{-1}$ and hence results in an infinite tower of Kaluza-Klein modes of mass scale $M_{KK}$ that cannot appear in QCD. Moreover, within this model, the phase transition between solitonic $D4$ branes and black $D4$ branes is interpreted as the strong coupling continuation of the confinement/deconfinement transition in 4 dimensional Yang Mills theory. This cannot be so, as the black $D4$ branes and the deconfinement phase of 4 dimensional Yang Mills have different realisations of the $Z_{N}$ centre symmetry and hence cannot be identified. This is why Mandal and Morita [10] proposed an alternative gravity dual for the confinement/deconfinement transition in terms of the Gregory-Laflamme transition of the soliton in the IIB theory, having as the strong coupling continuation of the deconfinement phase of the 4 dimensional Yang Mills a localised D3 soliton. Now, for the Sakai-Sugimoto model in the $D0-D4$ background, the smeared $D0$ charges added to the $D4$ background, as presented in ref. [41], have an important role in the understanding of topological changes in the system. The corresponding theory produces a glue condensate $\Bracket{tr(F_{\mu\nu}\tilde{F}^{\mu\nu})}$ and the $D8$ branes do go less deep than in the original Sakai-Sugimoto model. The goal of this paper is to understand a connection between phase transitions in gauge systems that may be physical, dual to gravitational models, and the extent to which quantum error correction can be naturally implemented in each of those systems' phases. As shown before, for this it is important to see how deep the gravitational dual objects enter in the bulk space and how that changes the capacity for quantum error correction. This may become useful in understanding what condensed matter systems and phases are best suited for quantum error correcting capable quantum computers.
Considering a $D0-D4$ background in which $D0$ branes have been added to the $D4$ background has a solution in type $IIA$ supergravity in Einstein frame as 
\begin{widetext}
\begin{equation}
\begin{array}{c}
ds^{2}=H_{4}^{-\frac{3}{8}}(-H_{0}^{-\frac{7}{8}}f(U)d\tau^{2}+H_{0}^{\frac{1}{8}}((dx^{0})^{2}+(dx^{1})^{2}+...+(dx^{3})^{2}))+H_{4}^{\frac{5}{8}}H_{0}^{\frac{1}{8}}(\frac{dU^{2}}{f(U)}+U^{2}d\Omega_{4}^{2})\\
\\
e^{-(\Phi-\Phi_{0})}=(\frac{H_{4}}{H_{0}^{3}})^{\frac{1}{4}}\\
\\
f_{2}=\frac{A}{U^{4}}\frac{1}{H_{0}}^{2}dU\wedge d\tau\\
\\
f_{4}=B\epsilon_{4}\\
\end{array}
\end{equation}
\end{widetext}
where 
\begin{equation}
\begin{array}{cc}
A=\frac{(2\pi l_{s})^{7}g_{s}N_{0}}{\omega_{4}V^{4}}, &
B=\frac{(2\pi l_{s})^{3}N_{c}g_{s}}{\omega_{4}}\\
H_{4}=1+\frac{U^{3}_{Q4}}{U_{3}}, &
H_{0}=1+\frac{U^{3}_{Q0}}{U^{3}}\\
\end{array}
\end{equation}
\begin{equation}
f(U)=1-\frac{U_{KK}^{3}}{U^{3}}
\end{equation}
with $d\Omega_{4}$, $\epsilon_{4}$ and $\omega_{4}=\frac{8\pi^{2}}{3}$ are the line element, the volume form, and the volume of a unit $S^{4}$ respectively. The coordinate radius of the horizon is $U_{KK}$, $V_{4}$ is the volume of the $D4$ brane. The number of $D0$ and $D4$ branes is $N_{0}$ and $N_{c}$. The $D0$ branes are smeared in the $x^{0},...,x^{3}$ directions. In the string frame this becomes 
\begin{widetext}
\begin{equation}
ds^{2}=H_{4}^{-\frac{1}{2}}(-H_{0}^{-\frac{1}{2}}f(U) d\tau^{2}+H_{0}^{\frac{1}{2}}dx^{2})+H_{4}^{\frac{1}{2}}H_{0}^{\frac{1}{2}}(\frac{dU^{2}}{f(U)}+U^{2}d\Omega_{4}^{2})
\end{equation}
\end{widetext}
where the euclidean form is used
\begin{equation}
dx^{2}=(dx^{0})^{2}+(dx^{1})^{2}+...+(dx^{3})^{2}
\end{equation}
and the equations of motion demand

\begin{equation}
\begin{array}{cc}
A^{2}=9 U_{Q0}^{3}(U_{Q0}^{3}+U_{KK}^{3}), & B^{2}9U_{Q4}^{3}(U_{Q4}^{3}+U_{KK}^{3})
\end{array}
\end{equation}

which are solved as
\begin{widetext}
\begin{equation}
\begin{array}{cc}
U_{Q0}^{3}=\frac{1}{2}(-U_{KK}^{3}+\sqrt{U_{KK}^{6}+\frac{4}{9}A^{2}}), &
U^{3}_{Q4}=\frac{1}{2}(-U_{KK}^{3}+\sqrt{U_{KK}^{6}+\frac{4}{9}B^{2}})
\end{array}
\end{equation}
\end{widetext}
After a wick rotation in the $\tau$ and $x^{0}$ directions the metric becomes 
\begin{widetext}
\begin{equation}
ds^{2}=H_{4}^{-\frac{1}{2}}(H_{0}^{-\frac{1}{2}}f(U)d\tau^{2}+H_{0}^{\frac{1}{2}}dx^{2})+H_{4}^{\frac{1}{2}}H_{0}^{\frac{1}{2}}(\frac{dU^{2}}{f(U)}+U^{2}d\Omega_{4}^{2})
\end{equation}
\end{widetext}
where now we have
\begin{equation}
dx^{2}=-(dx^{0})^{2}+(dx^{1})^{2}+...+(dx^{3})^{2}
\end{equation}
The metric is a bubble geometry and spacetime ends at $U=U_{KK}$. To avoid the conical singularity we can choose the period of $\tau$ as
\begin{equation}
\beta=\frac{4\pi}{3}U_{KK}H_{0}^{\frac{1}{2}}(U_{KK})H_{4}^{\frac{1}{2}}(U_{KK})
\end{equation}
The Kaluza Klein mass scale $M_{KK}=\frac{2\pi}{\beta}$ defines the UV cut-off in the gauge theory and the $D4$ brane tension is related to the five-dimensional Yang-Mills coupling constant
\begin{equation}
\frac{1}{g_{5}^{2}}=\frac{1}{(2\pi)^{2}l_{s}g_{s}}
\end{equation}
and using dimensional reduction to four dimensions we obtain the four dimensional Yang-Mills coupling constant
\begin{equation}
\frac{1}{g_{YM}^{2}}=\frac{\beta}{g_{5}^{2}}
\end{equation}
and reversely, we can express the string coupling in terms of gauge theoretical parameters as
\begin{equation}
g_{s}=\frac{g_{YM}^{2}}{2\pi M_{KK}l_{s}}=\frac{\lambda}{2\pi M_{KK}N_{c}l_{s}}
\end{equation}
with $\lambda=g_{YM}^{2}N_{c}$ which results into having 
\begin{equation}
\begin{array}{cc}
H_{0}(U_{KK})=\frac{1}{2}(1+(1+C\beta^{2})^{1/2}),& C=(2\pi l_{s}^{2})^{6}\lambda^{2}\tilde{\kappa}^{2}/U^{6}_{KK}
\end{array}
\end{equation}
In order to take into account the backreaction of the $D0$ brane we need $N_{0}\sim N_{c}$ and we can define 
\begin{equation}
\tilde{\kappa}=\frac{N_{0}}{(N_{c}V_{4})}
\end{equation}
which is what supplements this model when a $D0$ brane is introduced. 
and in the limit near the horizon with $U/\alpha'$ and $U_{KK}/\alpha'$ finite we obtain 
\begin{widetext}
\begin{equation}
\begin{array}{cc}
U_{Q4}^{3}\rightarrow \pi\alpha'^{3/2}g_{s}N_{c}=\frac{\beta g_{YM}^{2}N_{c}l_{s}^{2}}{4\pi}=R^{3}, &
H_{4}(U_{KK})\rightarrow \frac{R^{3}}{U^{3}_{KK}}\\
\\
\beta\rightarrow \frac{4\pi}{3}U^{-1/2}_{KK}R^{3/2}H_{0}^{1/2}(U_{KK}), &
M_{KK}\rightarrow \frac{3}{2}U_{KK}^{1/2}R^{-3/2}H_{0}^{1/2}(U_{KK})\\

\end{array}
\end{equation}
\end{widetext}
which brings us to the metric 
\begin{widetext}
\begin{equation}
\begin{array}{c}
ds^{2}=(\frac{U}{R})^{3/2}(H_{0}^{1/2}(U)\eta_{\mu\nu}dx^{\mu}dx^{\nu}+H_{0}^{-1/2}(U)f(U)d\tau^{2})+H_{0}^{1/2}(\frac{R}{U})^{3/2}(\frac{1}{f(U)}dU^{2}+U^{2}d\Omega_{4}^{2})\\
\\
e^{\Phi}=g_{s}(\frac{U}{R})^{3/4}H_{0}^{3/4}\\
\end{array}
\end{equation}
\end{widetext}
Therefore in the near horizon limit we obtain 
\begin{equation}
\beta^{1/2}=\frac{2}{3}\pi^{1/2}U_{KK}^{-1/2}\lambda^{1/2}l_{s}H_{0}^{1/2}(U_{KK})
\end{equation}
and
\begin{equation}
\begin{array}{cc}
\beta=\frac{4\pi\lambda l_{s}^{2}}{9U_{KK}}H_{0}(U_{KK}), &\;\; M_{KK}=\frac{9}{2}\frac{U_{KK}}{\lambda l_{s}^{2}H_{0}(U_{KK})}\\
\end{array}
\end{equation}
We can solve for $\beta$ 
\begin{equation}
\begin{array}{c}
\beta=\frac{4\pi\lambda l_{s}^{2}}{9U_{KK}}\frac{1}{1-\frac{(2\pi l_{s}^{2})^{8}}{81 U_{KK}^{8}}\lambda^{4}\tilde{\kappa}^{2}}\\
\\
H_{0}(U_{KK})=\frac{1}{1-\frac{(2\pi l_{s}^{2})^{8}}{81 U_{KK}^{8}}\lambda^{4}\tilde{\kappa}^{2}}\\
\end{array}
\end{equation}
Such a background introduces $\tilde{\kappa}$ as a new free parameter and therefore it is not dual to the vacuum state of the gauge theory. The dual state instead describes an excited state with a constant homogeneous field strength background which produces the expectation value of the term 
\begin{equation}
tr(F_{\mu\nu}\tilde{F}^{\mu\nu})
\end{equation}
On the gravity side, $\tilde{\kappa}N_{c}$ is the flux of $f_{2}$ and since $C_{1}$ is coupled to the term $tr(F_{\mu\nu}\tilde{F}^{\mu\nu})$ in the Euclidean Chern-Simons action, we see that $\tilde{\kappa}$ characterises the expectation value of the Euclidean $tr(F_{\mu\nu}\tilde{F}^{\mu\nu})$. We obtain therefore the real Euclidean condensate
\begin{equation}
\Bracket{tr(F_{\mu\nu}\tilde{F}^{\mu\nu})}=8\pi^{2}N_{c}\tilde{\kappa}
\end{equation}
With this we can go to the Sakai-Sugimoto model in $D0-D4$ backgrounds and we write, using $U=U(\tau)$ 
\begin{widetext}
\begin{equation}
\begin{array}{c}
ds^{2}=(\frac{U}{R})^{3/2}H_{0}(U)^{-1/2}(f(U)+(\frac{R}{U})^{3}\frac{H_{0}(U)}{f(U)}U'^{2})d\tau^{2}+\\
\\
+(\frac{U}{R})^{3/2}H_{0}^{1/2}(U)\eta_{\mu\nu}dx^{\mu}dx^{\nu}+H_{0}^{1/2}(\frac{R}{U})^{3/2}U^{2}d\Omega_{4}^{2}\\
\end{array}
\end{equation}
\end{widetext}
with $U'=\frac{dU}{d\tau}$. If we replace this into the $D8$ brane action we obtain 
\begin{equation}
S_{D8}\sim \frac{1}{g_{s}}\int d^{4}x d\tau H_{0}(U)U^{4}(f(U)+\frac{H_{0}(U)}{f(U)}(\frac{R}{U})^{3}U'^{2})^{1/2}
\end{equation}
which leads to the equation of motion
\begin{equation}
\frac{d}{d\tau}(\frac{H_{0}(U)U^{4}f(U)}{[f(U)+\frac{H_{0}(U)}{f(U)}(\frac{R}{U})^{3}U'^{2}]^{1/2}})=0
\end{equation}
and this can be solved for initial conditions $U(0)=U_{0}$, $U'(0)=0$ at $\tau=0$
\begin{widetext}
\begin{equation}
\tau(U)=E(U_{0})\int_{U_{0}}^{U}dU\frac{H_{0}^{1/2}(U)(\frac{R}{U})^{3/2}}{f(U)(H_{0}^{2}(U)U^{8}f(U)-E^{2}(U_{0}))^{1/2}}
\end{equation}
\end{widetext}
with $E(U_{0})=H(U_{0})U_{0}^{4}f^{1/2}(U_{0})$. We can see that as opposed to the $D4$ soliton background, the $D0-D4$ background contains $H_{0}$ factors in all equations. We obtain back the original Sakai-Sugimoto model if we set $H_{0}(U)\rightarrow 1$. As the $D8-\bar{D8}$ moves away from the antipodes we can see that the evolution is less deep inside the bulk as opposed to the original model. Therefore it is important to see the connection between the topological term and the addition of $D0$ branes into the system and the evolution of the brane system within the bulk space. As $\tilde{\kappa}$ grows away from zero, the profile of the D8 brane corresponds to a less deep configuration. This is linked to a lower capacity for quantum error correction. 
The example above follows closely the results obtained in ref. [41], mainly the introduction of the $\tilde{\kappa}$ parameter due to the smeared $D0$ charges. Even in this model we can see that a glue condensate term reduces the capacity for quantum error correction due to the fact that it stops the profile of the $D8$ brane reaching deep inside the bulk. 

In the case of the $D0-D4$ system we distinguish two situations. One in which the $D0$ brane moves in a black brane $D4$ geometry, case in which the $D0$ brane is not affected by the distance the $D4$ branes reach inside the bulk, and one in which the $D0$ brane moves on a solitonic $D4$ geometry. As we move along the radial bulk coordinate inside the bulk of the solitonic $D4$ geometry, the size of the instanton diminishes as it gets closer to the tip of the D4-soliton. This solitonic tip lies at the deepest accessible position inside the bulk and is obviously related to the instanton size. From a QCD instanton dynamics point of view, the position $u$ of the D0-brane would be related to the size $\rho$ of the instanton by the relation [9]
\begin{equation}
\rho\sim\frac{\sqrt{\lambda_{p}}}{u^{(5-p)/2}}
\end{equation}
where $\lambda_{p}$ is the 't Hooft coupling on the Dp-branes. In the confinement geometry the existence of two energy scales makes such a formula difficult to apply, but we can accept the assumption of ref. [9] that 
\begin{equation}
\rho\sim\sqrt{\frac{L_{4}\lambda_{YM}}{u}}
\end{equation}
where $\lambda_{YM}$ is the dimensionless coupling $\lambda_{YM}=\frac{2\lambda_{5}}{L_{4}}$ and $\rho$ is the size of a QCD instanton. 
This assumption is valid at least in the $u\gg u_{0}$ limit. The DBI action [15] shows the suppression of the small instantons in the confinement phase, a characteristic consistent with the results of perturbative QCD. In the region where $u\sim u_{0}$ and $\rho\sim\sqrt{L_{4}\lambda_{YM}/u_{0}}\sim L_{4}$ we can have an instanton with zero value of the DBI action, meaning we have strong fluctuations of the topological charge. However, larger instantons cannot exist and hence the instanton density will have a sharp peak at $\rho_{peak}\sim L_{4}$ for large $N$. The description in terms of a D4-brane system and a D0-instanton already shows that in the directions given by the $(x^{5},...,x^{9})$ coordinates there is a region deep within the bulk that is not accessible characterised by $u_{0}$. Let us hence cover this space with a tensor network according to the prescriptions of the holographic quantum error correction codes. Starting on the boundary which lies in the region $u\rightarrow \infty$ we can follow the links towards the interior of the bulk on the D4-brane described above. The tip of the D4-soliton represents a maximal region that can be reached inside the bulk in the confinement phase, and such a limit translates into a limit to the quantum error correction capabilities of a holographic quantum code using a confined QCD phase as its substrate. 

To describe the QCD thermodynamics in the high temperature regime and hence to move towards the deconfinement transition, one has to consider the T-dual along the Euclidean time circle [9]. The system will then be constructed out of N D3-branes and the QCD instanton will be represented by a D1-brane instead of a D0-brane moving on the D3-brane background. Defining $T_{c}$ the critical temperature, in the domain $T<T_{c}$ the stable geometry in the T-dual frame is given by the uniformly smeared D3-branes corresponding to the solitonic D4-brane geometry in the previous context. The instanton is described by a D1-brane on this geometry. At temperatures higher than the critical temperature, the stable geometry is identified with the localised D3-branes. The dynamics of the D1-branes in this geometry will give us insights on the instanton effects. The deconfinement phase therefore corresponds to a localised solitonic D3-brane solution on the $t'$-cycle which corresponds to the Euclidean time direction in the T-dual frame. The location of the centre of mass of these D3-branes can be considered at $t'=0$ and due to periodicity $t'=t'+\beta'$ where $\beta'=4\pi^{2}T$. Their mirror images are placed around $t'=n\beta'$ with $n=\pm1,\pm2,...$. The system constructed out of such branes has interacting components, the D3-brane and its mirror being gravitationally coupled and hence. This will lead to the breaking of the spherical symmetry. The corrections due to non-sphericity can however be ignored in a first approximation. Following ref. [9] as well as [16, 17] we may consider the Wick rotation of the black D3 brane localised on the $S^{1}$ circle and we take over the metric in the context of our solitonic branes
\begin{widetext}
\begin{equation}
ds^{2}=H^{-1/2}[\sum_{i=0}^{3}dx_{i}^{2}+fdx_{4}^{2}]+H^{1/2}[\frac{A}{f}dR^{2}+\frac{A}{K^{d-2}}dv^{2}+KR^{2}d\Omega_{4}^{2}]
\end{equation}
\end{widetext}
with $f=1-\frac{R_{0}^{3}}{R^{3}}$
where $A$ and $K$ are functions of $R$ and $v$ and $x^{4}$ is the direction for $S_{L_{4}}^{1}$ before we take the T-dual on the time direction. The Euclidean time coordinate $t'$ is included in the $(R,v)$ plane. 
This metric can be divided into the asymptotic region and the near region, with the asymptotic region being characterised by $u\gg u_{H}$ or $t'\gg u_{H}$. Here the effects of the black hole are small and can be calculated by linearising the equations resulting in the effective metric 
\begin{widetext}
\begin{equation}
ds^{2}=\alpha'[H^{-1/2}(\sum_{i=1}^{3}dx_{i}^{2}+(1+2\Phi)dx_{4}^{2})+H^{1/2}(1-\frac{1}{2}\Phi)(du^{2}+dt'^{2}+u^{2}d\Omega_{4}^{2})]
\end{equation}

\begin{equation}
\begin{array}{cc}
H=\sum\limits_{n}\frac{2\lambda_{5}/\beta}{(u^{2}+(t'-n\beta')^{2})^{2}}, & e^{\phi}=\frac{\lambda_{5}}{2\pi N\beta}\\
\end{array}
\end{equation}

\begin{equation}
\begin{array}{ccc}
\Phi=-\frac{u_{H}^{4}}{2}\sum\limits_{n}(\frac{1}{u^{2}+(t'-n\beta')^{2}})^{2}, & u_{H}=\sqrt{2\lambda_{5}T}\frac{\pi}{2L_{4}}, & \beta'=\frac{(2\pi)^{2}}{\beta}=(2\pi)^{2}T\\
\end{array}
\end{equation}
\end{widetext}
This represents the asymptotic limit where the mirror images contribute to the metric. 
The metric in this region is given by [9]

\begin{widetext}
\begin{equation}
ds^{2}=\alpha'[H^{-1/2}(\sum\limits_{i=1}^{3}dx_{i}^{2}+(\frac{1-\frac{r_{0}^{4}}{r^{4}}}{1+\frac{r_{0}^{4}}{r^{4}}})^{2}dx_{4}^{2})+H^{1/2}(1+\frac{r_{0}^{4}}{r^{4}})(dr^{2}+r^{2}d\Omega_{5}^{2})]
\end{equation}
\end{widetext}
\begin{widetext}
\begin{equation}
\begin{array}{ccc}
H=\frac{2\lambda_{5}/\beta}{r^{4}}(1+\frac{r_{0}^{4}}{r^{4}})^{-2}, & r_{0}=\frac{u_{H}}{\sqrt{2}}=\frac{\pi\sqrt{\lambda_{5}T}}{2L_{4}}, & e^{\phi}=\frac{\lambda_{5}}{2\pi N\beta}
\end{array}
\end{equation}
\end{widetext}
The instanton described by the D1-brane geometry in the asymptotic region can be seen as embedded in the $(t',x_{4},u)$ space and $(t',x_{4})$ can be taken as the worldvolume coordinates on the D1-brane. The induced metric is then

\begin{equation}
ds_{D1}^{2}=\alpha' [H^{-1/2}(1+2\Phi)dx_{4}^{2}+H^{1/2}(1-\frac{1}{2}\Phi)(1+(\frac{dU(t')}{dt'})^{2})dt'^{2}]
\end{equation}

and then the DBI action in this region becomes 
\begin{widetext}
\begin{equation}
\begin{array}{c}
S_{D1}=\frac{1}{(2\pi)\alpha'}\int\limits_{0}^{\frac{(2\pi)^{2}}{\beta}}dt'\int\limits_{0}^{L_{4}}dx_{4}e^{-\phi}\sqrt{det(g_{D1})}=\\
=\frac{N\beta L_{4}}{\lambda_{5}}\int\limits_{0}^{\frac{(2\pi)^{2}}{\beta}}dt'(1+\frac{3}{4}\Phi)\sqrt{1+(\frac{dU(t')}{dt'})^{2}}\\
\end{array}
\end{equation}
\end{widetext}
As has been calculated in [9], the D1-brane instanton is attracted towards the D3-branes at $u=0$ as in the confinement case. This result however breaks down as $u\sim u_{H}$. As in the deconfined phase the geometry allows the position $u=0$ and the D1-brane is attracted towards the D3-brane, it would stabilise at $u=0$ and would stretch between the D3-brane and its mirror image along the $t'$ compact circle. At this point, the localised D3-brane has the topology $S^{2}\times R^{3}\times S^{4}$ with the stable D1-instanton brane warping around the $S^{2}$. In the asymptotic region however, where $u\gg u_{H}$ the D1-brane would warp around the $t'$ and $x_{4}$ cycles composing the topology of a 2-torus $T^{2}$. This shows that when the D1-brane reaches the D3-brane horizon, its topology must change and hence the small instanton cannot continuously transform into the stable instanton. 
The main idea is to connect the phase structure in gauge theory with some equivalent relation on the higher dimensional gravity side. This connection has historically created many problems. A naive interpretation of holography to a two dimensional large N bosonic gauge theory at finite temperatures would imply a phase diagram on the gravity side that would not admit a unique continuation to the gauge theory side because of its dependence on the boundary condition for fermions on the brane. Such phase structures have however been shown to be smoothly connectable by making appropriate choices of fermion boundary conditions. If for example an anti-periodic boundary condition is being taken for the fermions on the brane, the fermions will get masses (for example proportional to the Kaluza Klein scale in case of an $S^{1}$ spacetime extension), resulting in supersymmetry breaking. It has been shown in [10] that the confinement/deconfinement transition may correspond to a Gregory-Laflamme transition between a uniformly distributed D3 soliton and a localised D3 soliton in the stringy IIB frame. 
It must be mentioned that the construction of ref. [9] is incomplete and becomes invalid once one observes that the black D4 brane and the deconfinement phase of YM4 have different realisations of the $Z_{N}$ symmetry. On one side this would simplify the construction of the effective metric in the two cases as no horizons will appear, however, it is visible that the same problem will appear there as well: certain deep bulk regions will not be reachable in the case of the confinement phase leading to a lower level of quantum error correction. While there is no black hole or black brane horizon in the accurate configuration of ref. [10], the geometry of the D-brane inside the bulk is constructed such that following the coordinates inside the bulk one reaches the top of the brane hence any object connected to that brane will not have access deep inside the bulk. The other situation however allows for a smooth continuation deeper in the bulk. As can be seen in the pictures the directions $x_{1}$, $x_{2}$, $x_{3}$ are intrinsic and common to the D-branes, the temporal directions are periodic, and the directions $x_{5},...,x_{9}$ are being covered by the D-branes in the deconfined configurations while in the confined configurations the geometry of the D-brane is being pinched off. This makes our D-brane in this configuration unable to extend arbitrarily in the $x_{5},...,x_{9}$ directions. In fact the same argument is being brought in ref. [10] for the characterisation of chiral symmetry breaking in a $D4-D8/\bar{D8}$ configuration and it would be interesting to see how the pinching off of this geometry will limit quantum error correction capabilities in the case of a chiral-non-chiral symmetry phase transition. 
Ref. [10] proposes a reinterpretation of the confinement/deconfinement phase transition in terms of a Gregory-Laflamme transition between the solitonic D4 brane and the T-dual of the localised solitonic D3 brane. Studying the entanglement entropy via a Ryu Takayanagi approach in this case will have to take into account the compatibility between the geometry in the gauge limit and the geometry of the higher dimensional gravitational theory. This will require a covariant Ryu Takayanagi approach and will have to take into account the T-dual brane structure appearing in the process. The T-duality along the temporal direction in the bulk D3-brane geometry makes the matching of the bulk temporal direction and the boundary temporal direction not equivalent.

\section{Quantum error correction and phase transition}
In the previous chapter I showed that the geometry and topology of the confined and deconfined phases of QCD change in the process of phase transition and regions deep in the bulk become non-trivially connected to regions closer to the boundary when the transition towards a deconfined phase occurs. As a result a tensor network will detect inaccessible regions inside the bulk leading to limitations to its error correction capabilities.  We have seen that in the confined phase the geometry of the solitonic D4-branes cannot extend indefinitely in the $(x^{5},...,x^{9})$ directions as it is being pinched off at $u=u_{0}$. If a tensor network is to be constructed on such a structure starting close to the boundary in the large $u\rightarrow \infty$ region, then several regions of the bulk will not be accessible. As will be shown in this chapter, this limits the quantum error correction codes that can be constructed on this confining geometry. On the other side, for the deconfinement phase, the horizontal bulk directions $(x^{5},...,x^{9})$ are not totally pinched off, and instead, the D3-branes will be connected by means of a topologically non-trivial link. If one constructs a tensor network from the boundary towards the bulk, one will not face any limit in reaching the maximal error correction potential deep inside the bulk. 
Let us therefore start with the construction of the tensor network and the holographic code for these two configurations. 
To construct tensor network representations and to make meaningful use of the MERA (multi-scale entanglement renormalisation ansatz) [18] in a general space the surface/state correspondence is particularly useful [19]. In this general context one can initiate the construction of a tensor network without the need of an explicit boundary. While certainly the AdS/QCD correspondence implies the existence of a boundary that would be associated to the QCD theory, the particularities of the bulk may require some general techniques corresponding to the surface/state duality principle [19]. It has been shown in [10] that the high temperature region in gravity corresponds to a localised solitonic D3 brane. Its centre symmetry matches with the deconfinement phase of four dimensional Yang Mills theory. This brane structure replaces our bulk space which corresponds now to QCD and hence the transition occurs between a confined and deconfined phase for the D3-D1 system. We will be interested in determining the quantum information properties of the brane structure inside the bulk and derive from it a connection between phase transition and quantum error correction capabilities in the bulk space. It is assumed that a tensor network structure can be constructed on the D3-D1 system. While the structure of such a network will be different, the meaning will be the same as the one employed in AdS/CFT i.e. it will result in a quantum field theory with manifest entanglement on the boundary and will implement quantum error correction protocols against erasures in the bulk. Indeed, in this context we define the infinite dimensional Hilbert space associated to the bulk spacetime as $\mathcal{H}_{tot}$ and we focus on a subset of this space given by a convex surface $\Sigma$. If the surface $\Sigma$ is topologically trivial and hence homological to a point, the associated quantum state is a pure state $\ket{\Phi(\Sigma)}\in\mathcal{H}_{tot}$. When the surface $\Sigma$ is topologically non-trivial, the corresponding state becomes mixed and the associated Hilbert space becomes a subspace of $\mathcal{H}_{tot}$ so that 
\begin{equation}
\rho(\Sigma)\in End(\mathcal{H}_{\Sigma})\leftrightarrow \Sigma \in M
\end{equation}
The subspace $\mathcal{H}_{\Sigma}$ only depends on the topological class of $\Sigma$. This plays an important role in the context of confinement-deconfinement phase transition when there is a relevant variation of the topological charge. If the surface $\Sigma$ is not closed, but instead has a boundary $\partial\Sigma$ the dual quantum state becomes mixed again, and if $\Sigma$ can be associated to a submanifold of another convex surface $\tilde{\Sigma}$ that is closed then the mixed state is given by tracing over the Hilbert space corresponding to the complement of $\Sigma$ in $\tilde{\Sigma}$. When we have two surfaces $\Sigma_{1}$ and $\Sigma_{2}$ that are connected by a smooth deformation preserving convexity, then this deformation in the quantum dual state can be described by means of an integral of infinitesimal unitary transformations
\begin{equation}
\begin{array}{c}
\ket{\Phi(\Sigma_{1})}=U(s_{1},s_{2})\ket{\Phi(\Sigma_{2})}\\
U(s_{1},s_{2})=P\cdot exp[-i\int\limits_{s_{2}}^{s_{1}}\hat{M}(s)ds]\\
\end{array}
\end{equation}
with $P$ being the path ordering operator and $\hat{M}(s)$ is a hermitian operator where the parameter $s$ describes the continuous deformation such that its extremes $s_{1}$ and $s_{2}$ correspond to the two surfaces. To apply MERA on the brane structure describing the confinement and the deconfinement phases we need to keep under control the evolution of the number of links in the tensor representation that intersect the surface we want to consider towards the rest of the brane, such that it covers the whole accessible regions on the brane structure. Even though in the surface/state duality the area $\Sigma$ is not an extremal surface in the most general case, it is always possible to divide it in small enough subregions that can be considered as extremal [19]. The geometry of such a subregion is almost flat and hence the extremal surfaces become themselves flat planes. Summing up such flat surfaces and accepting the Ryu-Takayanaga prescription [20], [21] we obtain 
\begin{equation}
\sum_{i}S_{A_{i}}^{\Sigma}=\frac{A(\Sigma)}{4G_{N}}
\end{equation}
where $A_{i}$ corresponds to the small portion of $\Sigma$ defined as $\Sigma=\cup_{i}A_{i}$ with the intersection of the small pieces being the void set. The left side of the equation above is the effective entropy $S_{eff}(\Sigma)$ [19] and is being interpreted as the logarithm of the effective dimension of the Hilbert space $\mathcal{H}_{\Sigma}$ namely 
\begin{equation}
S_{eff}(\Sigma)=log[dim(\mathcal{H}_{\Sigma}^{eff})]
\end{equation}
The effective dimension counts the dimension of the effective degrees of freedom corresponding to the entanglement between $A_{i}$ and its complement and therefore $S_{eff}(\Sigma)$ is of the order of magnitude of the number of links in the tensor network intersecting our surface $\Sigma$ as required. The MERA scheme implies a real-space renormalisation based on the coarse-graining of the original system by combining step by step two subsystems into a single subsystem according to a linear map (also known as isometry). The short range entanglement however must be taken into account and cut-off after each coarse graining process by means of a unitary transformation called disentangler. The final state hence becomes a network of substates which, in its tensorial form spans the branes in each of the phases. In the AdS/CFT duality, such a MERA network can be identified with the bulk AdS spacetime. In our context however, we may start a similar procedure close to the boundary, following the soliton D3 brane towards the bulk. In the confinement phase, we can consider a set of quantum states $\ket{\Psi(u)}$ where $u$ refers to a bulk coordinate. The vacuum state $\ket{\Psi(0)}$ can be considered as being highly entangled and hence it can be constructed from the trivial state $\Omega$ associated to the state $\Sigma$ degenerating to a point, by adding to it entanglement by means of 
\begin{equation}
\ket{\Psi(u)}=P\cdot e^{-i\int\limits_{u_{IR}}^{u}(K(s)+L)ds}\ket{\Omega}
\end{equation}
The Hermitian operators $K(s)$ and $L$ describe the disentangler and coarse graining procedure. In terms of general states, one can write 
\begin{equation}
\ket{\Phi(u)}=e^{iuL}\ket{\Psi(u)}=P\cdot e^{-i\int\limits_{u_{IR}}^{u}\hat{K}(s)ds}\ket{\Omega}
\end{equation}
where $\hat{K}(s)$ is the disentangler in the interaction picture
\begin{equation}
\hat{K}(u)=e^{iuL}K(u)e^{-iuL}
\end{equation}
After disentangling for the states $\ket{\Psi(u)}$ we also perform a rescaling, while this procedure is absent for $\ket{\Phi(u)}$. 
The particularity here is that the procedure must also consider the additional links in the direction of the D1-brane associated to the instanton. Indeed, this will contribute to the effective dimension of our Hilbert space and hence to the effective entropy 
\begin{equation}
S_{eff}(\Sigma)=log[dim(\mathcal{H}_{\Sigma}^{eff}\otimes D1)]
\end{equation}
The additional links can be seen in figure 1. 
\\
\begin{figure}
  \includegraphics[width=180pt]{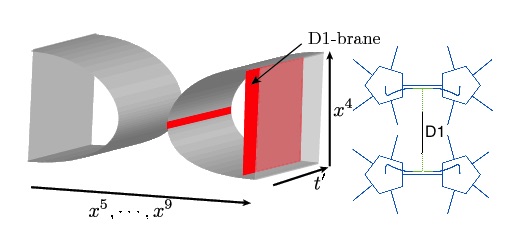}
  \caption{The tensor network construction on the D3 brane with additional links in the direction covered by the D1 instanton. The links in the D1 direction will stop at the tip of the D3 brane. Each link in the figure may have additional components according to the dimensionality of the subspace it points to}
  \label{fig:fig1}
\end{figure}
\\

The quantum information description involves both the theory on the boundary and the bulk geometry. While in the bulk the D3-D1 configuration limits the amount of entanglement, when we measure via Ryu Takayanagi (the area of the geometry to be analysed within the bulk depends on the D3-D1 geometry) in order to obtain the equivalent for the boundary theory, similar limitations will occur. It is therefore expected to have quantum information effects occurring both in the bulk and on the boundary and to be linked via dual relations. For example the stability to quantum errors on the boundary should be recoverable by the depth of the tensor network inside the bulk. When a limitation on the spreading of the tensor network inside the bulk is imposed, a limitation on the resistance to quantum errors in the boundary is to be expected.

As we continue the procedure towards $u=0$ we notice that due to the configuration of the branes in the confinement phase, it is impossible to reach certain regions of the bulk, those regions acting like barriers that do not allow links of our networks to cross them. This is equivalent to saying that error correction algorithms based on the bulk encoding of information cannot reach the highest quantum error correction capability of which the holographic encoding is usually capable. This is easy to see if one considers the tip of the D3 soliton at $u_{0}$ and follows the MERA prescription outwards towards the boundary. If one remains on the D3 brane as required within the bulk, then one notices that the region of the boundary that must be deleted in order to negatively affect the information encoded near the tip of the D3 brane is encoded on a smaller region of the boundary compared to the normal holographic tensor network case. Now let us consider the deconfinement case. From a quantum information point of view, we notice that in this case, the topology has become that of $S^{2}\times R^{3}\times S^{4}$ and the instanton D1-brane wraps around $S^{2}$ while in the asymptotic region, the D1-brane wraps around the $t'$ and $x_{4}$ cycles forming the topology of $T^{2}$. Therefore, as the D1 brane moves from the asymptotic to the near region we have a change in the topology. This is revealed in the fact that the associated surface will have a non-trivial topology to be associated with a mixed state. In this situation, the available quantum error correction becomes more powerful and the limitation due to the finite tip of the D3-brane is partially eliminated. However, one has to consider the limitations due to the fact that the state described by this structure is at this point a mixed state. A detailed study of quantum error correction codes for mixed states has been given in [22], [23], [37]. The actual situation is presented in figure 2. 
\\
\begin{figure}
  \includegraphics[width=180pt]{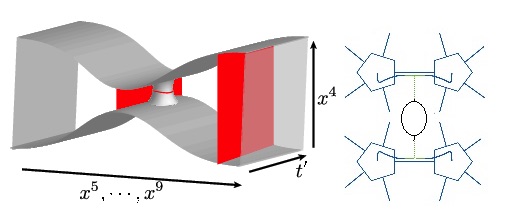}
  \caption{The throat forming in the deconfinement phase allows the links in the tensor network to penetrate. The situation is similar to that of a spacetime with a wormhole geometry.}
  \label{fig:fig1}
\end{figure}
\\
In this situation, the links can pass through the throat forming at $u=0$ and hence the quantum error correction code allows for a boundary encoding with more effective area. 
The explicit calculation of the entanglement entropy in this context is somehow more involved as it includes certain topological particularities specific to this problem. One relatively well known method of obtaining the entanglement entropy is by means of the so called replica trick. It has indeed been used both in perturbative as well as non-perturbative situations with remarkable success and hence could be in principle applied to the system made up of a D1 and a D3 brane. The D1 configuration can be seen as a one dimensional string evolving in time, linked to the D3 brane. 


The most important phenomenon arising here is a non-perturbative effect encoded by the appearance of a wormhole type geometry deep inside the bulk. That geometry adds a torus like topology to our problem. To properly apply the replica trick we have to consider a path integral over a $q$ sheeted surface obtained by gluing together $q$ replicas of our original surface along specially chosen cuts called generally entangling surfaces. Work with the replica trick has been done on related subjects in ref.  [34], [35], [36]. The situation here is being complicated by the presence of the non-trivial topology which makes the problem intrinsically non-perturbative. The method works from a practical point of view in the following way: one considers our surface $\mathcal{C}$ and defines a region $A$ on it at some fixed time. The reduced density matrix associated to the region $A$ can be computed by means of the usual trace formula
\begin{equation}
\rho_{A}=Tr_{B}\rho
\end{equation}
where $\rho$ is the full quantum state defined at a given time on $\mathcal{C}$ and $B$ represents the complement of $A$ with the trace being carried over the degrees of freedom from the complement of $A$. 
Explicitly the trace over the replica density matrix will be given by 
\begin{equation}
Tr(\rho^{q}_{A})=\frac{Z_{q}(A)}{(Z_{1}(A))^{q}}
\end{equation}
where $Z_{q}(A)$ is the path integral over the $q$-sheeted surface gluing together the $q$ replicas. Obviously the $Z_{1}$ path integral is the path integral over the single non-replicated surface. $A$ represents the cut across which the gluing occurs. Now the von Neumann entropy can be calculated by a set of interesting prescriptions. First one continues $q$ analytically into the continuous real numbers, then one differentiates with respect to it, and then one takes the limit $q \rightarrow 1$. 
This prescription is clearly mathematically controversial. First of all, $q$ is defined as a counting index for the number of replicated surfaces. What is the meaning of an analytic continuation for such a number? Are there any singularities to be considered in the process? Is the limit well defined? The replica trick in itself originates from the theory of disordered statistical systems where it has been extensively used and it has been checked to work in most practical cases given sufficient care in handling the commutation of various limits arising in these theories. Not insisting upon the unclear mathematical details at this point (despite them being intrinsically important) I can use the constructions of the replica trick to design a solution for the topologically non-trivial part of our D1-D3 brane system. The topologically disconnected one has been calculated before and I will cite authors like [24] for that. That contribution will here be simply denoted by $S_{dis}$. 
In the deconfinement phase we obtain a throat that in the neighbourhood of the contact to the brane behaves like an AdS space. The rest of it looks like a tube, which for all practical purposes here can be described as a cylinder. The question we have to ask therefore is what will be the entanglement entropy of a string when it gets into contact with such a cylinder. To apply the replica trick in this case amounts to gluing not only the D3 sheets but also the cylinders in order to obtain a consistent $q$-sheeted surface. Indeed, there will be $q$ such cylinders glued to each other by means of cuts which can be performed along the cylinder. The surfaces themselves will be trivially glued across the $q$ sheets. Tracing over $B$ in this case plays the role of tracing over the external configuration outside the cylinder. Our D1 brane (the string) is described by means of both localised oscillatory modes across itself, carrying momentum and extended topologically inequivalent winding numbers. The perturbative modes are easily calculated and not relevant now. The topological winding numbers need to be properly characterised in this context. 
Take therefore the case in which our surface $\mathcal{C}$ is a cylinder and $A$ is a cut going longitudinally along its length. In this case $Z_{1}$ is the usual cylinder amplitude and I will note it $Z$. The worldsheet has the topology of a cylinder and the entanglement surface is the cut covering the whole longitudinal length. 
We can consider here the target space to be a 2-dimensional torus times some non-compact spectator dimensions. The torus has two cycles parametrised by the vectors $R_{1}$ and $R_{2}e^{i\alpha}$, with $R_{1}, R_{2}, \alpha \in \mathbb{R}$. The complex structure will be given by $\sigma=\frac{R_{2}}{R_{1}}e^{i\alpha}$. Using the target space coordinates $x^{\alpha}=(x^{1}, x^{2})$ we can write the complex coordinates $ (z, \bar{z}) $ making use of $z=R_{1}x^{1}+R_{2}e^{i\alpha}x^{2}$. 
The replica trick for the entanglement entropy can be written as 
\begin{widetext}
\begin{equation}
S_{A}=-Tr(\rho_{A} log(\rho_{A}))=lim_{n\rightarrow 1} \frac{log(Tr \rho_{A}^{n})}{1-n}=-lim_{n\rightarrow 1}\frac{\partial}{\partial n} Tr \rho_{A}^{n}
\end{equation}
\end{widetext}
the argument of the first limit being the Renyi entropy. One can think in terms of a density matrix $\rho$ of a thermal state at temperature $T=1/\beta$ with the expression
\begin{widetext}
\begin{equation}
\rho(\{\phi_{x}\},\{\phi_{x'}'\})=\frac{1}{Z}\int d\phi(y,\tau)\prod_{x'}\delta(\phi(y,0)-\phi_{x'}')\prod_{x}\delta(\phi(y,\beta)-\phi_{x})e^{-S_{E}}
\end{equation}
\end{widetext}
with 
\begin{equation}
Z=Tr(e^{-\beta H})
\end{equation}
The role of the trace is to bind together the edges at $\tau = 0$ and $\tau = \beta$ producing a cylinder with circumference of length $\beta$. If we perform a partial trace 
\begin{equation}
\begin{array}{c}
\rho_{A}=Tr_{B}\rho\\
A=(u_{1},v_{1})\cup ... \cup (u_{N},v_{N})\\
\end{array}
\end{equation}
Here the union goes over all open cuts used to glue different sheets. $N$ will count these open cuts independently. The intervals $(u_{i}, v_{i})$ represent intervals forming the gluing sheets for the replica copies. 
In this case the trace only binds together the points not belonging to $A$. Open cuts are being left along the intervals $(u_{j},v_{j})$. These can in general be used to produce $q$ copies of cylinders bound together along the cuts. This leads us to 
\begin{equation}
Tr \rho_{A}^{q}=\frac{Z_{N,q}}{Z^{q}}
\end{equation}
The partition function on the $q$ sheeted surface $R_{N,q}$ will be given as
\begin{widetext}
\begin{equation}
\begin{array}{c}
Z_{N,q}=\int_{C_{A}}d\phi_{1}...d\phi_{q} exp[-\int dz d\bar{z}(\mathcal{L}[\phi_{1}](z,\bar{z})+...+\mathcal{L}[\phi_{q}](z,\bar{z}))]\\
\\
C_{A}: \phi_{i}(x,0^{+})=\phi_{i+1}(x,0^{-}), x\in A=\cup_{j=1}^{N}, i=1,...,q\\
\end{array}
\end{equation}
\end{widetext}
Our case is actually much simpler because the connection with the parallel branes will be done across the entire circumference of the cylinder at each of the two ends. This actually makes the calculation very similar to one that has already been performed in [24-26]. 

As said before we are mostly interested in the topological problem, with the observation that we need to pay attention to area changes in order to properly take into account the entanglement entropy, as it depends on the bulk area. Therefore our model theory will follow ref. [27] and will focus on a set of theories called Quasi-Topological Quantum Field Theories, or QTFT. Such theories are fundamentally topological quantum field theories with the observation that in a normal topological quantum field theory surfaces with different areas can belong to the same topological class. This is highly undesirable if our goal is to understand realistic entanglement entropies, as it is known that such entropies are linked to surface areas. Therefore, we will need to extend our theory towards a construction where different areas are considered independently. Quasi-topological quantum field theories do exactly this as their correlation functions depend on the topology and, additionally, on the total area of the underlying spacetime while being blind to other aspects of the geometry. The replica trick, involving the evaluation of path integral on replicated Riemann surfaces can therefore be used and meaningful results can emerge. Take therefore our surface as depicted in figure 2 and let us trace it along the imaginary time direction. This will effectively compactify our surface into a torus and will allow us to work with closed surfaces. Of course the compactification radius can always be set to infinity leading to our example from figure 2. By doing this we obtain the representation of figure 3. As there exist several generalisations to the Dp-D(p+2) branes coupled states [28] I will focus on a theory living on the D-brane that exists in two dimensions for pragmatic reasons. The Migdal formalism allows us to construct a theory in two dimensions invariant under area-preserving diffeomorphisms. In two dimensions it is easier to see how the cuts are being performed. Therefore, looking at figure 3, partial trace over the $B$ degrees of freedom allows us to construct our torus with a set of cuts which then can be used to glue the surface to our replicas. If we consider the two dimensional Yang Mills theory of the form
\begin{widetext}
\begin{equation}
S[A]=-\frac{1}{2e^{2}}\int_{\Sigma} Tr(F\wedge \star F)=-\frac{1}{4e^{2}}\int_{\Sigma}\sqrt{g}d^{2}x g^{\mu\sigma}g^{\nu\rho}Tr(F_{\mu\nu}F_{\sigma\rho})
\end{equation}
\end{widetext}
where $g(x)$ is the background metric of the Riemann surface $\Sigma$ and the curvature 2-form is constructed from the gauge field as $F=dA+A\wedge A$.
To define a gauge theory one has to identify a gauge group $G$ and to construct a principal bundle $P_{\Sigma, G}$ with fibres in this group. The $YM_{2}$ theory is quasi-topological. To see this note that the curvature is proportional to the volume form $\epsilon$ and is a gauge covariant 2-form. Hence one can write $F=f\cdot \epsilon$ and $f$ transforms to the adjoint representation of $G$. This leads us to 
\begin{equation}
S=-\frac{1}{2e^{2}}\int_{\Sigma} \epsilon Tr(f^{2})
\end{equation}
The explicit metric dependence has been replaced by the volume form as should be the case in a quasi-topological quantum field theory that keeps sensitive only to the area. This makes the $YM_{2}$ theory see areas but not distances. A continuum 2-dimensional Yang Mills theory can be reduced to an exactly solvable theory on a lattice. This two-dimensional theory is defined on our D-brane and corresponds to the boson sector of our description. This is indeed a mathematical simplification that allows us to easily (or at least less difficultly) calculate the entanglement entropy in a geometric setting. 
The formalism used for that is called the Migdal formalism, discussed in ref. [27]  and it leads us to the expression for the partition function
\begin{equation}
Z(\rho,g)=\sum_{R\in irrep(G)} d_{R}^{2-2g}e^{-\frac{e^{2}\rho c_{2}(R)}{2}}
\end{equation}
The sum is over all irreducible representations of the gauge group $G$, $d_{R}$ represents the dimension of the representation $R$ while $c_{2}(R)$ the quadratic Casimir of the representation $R$.
\\
\begin{figure}
  \includegraphics[width=120pt]{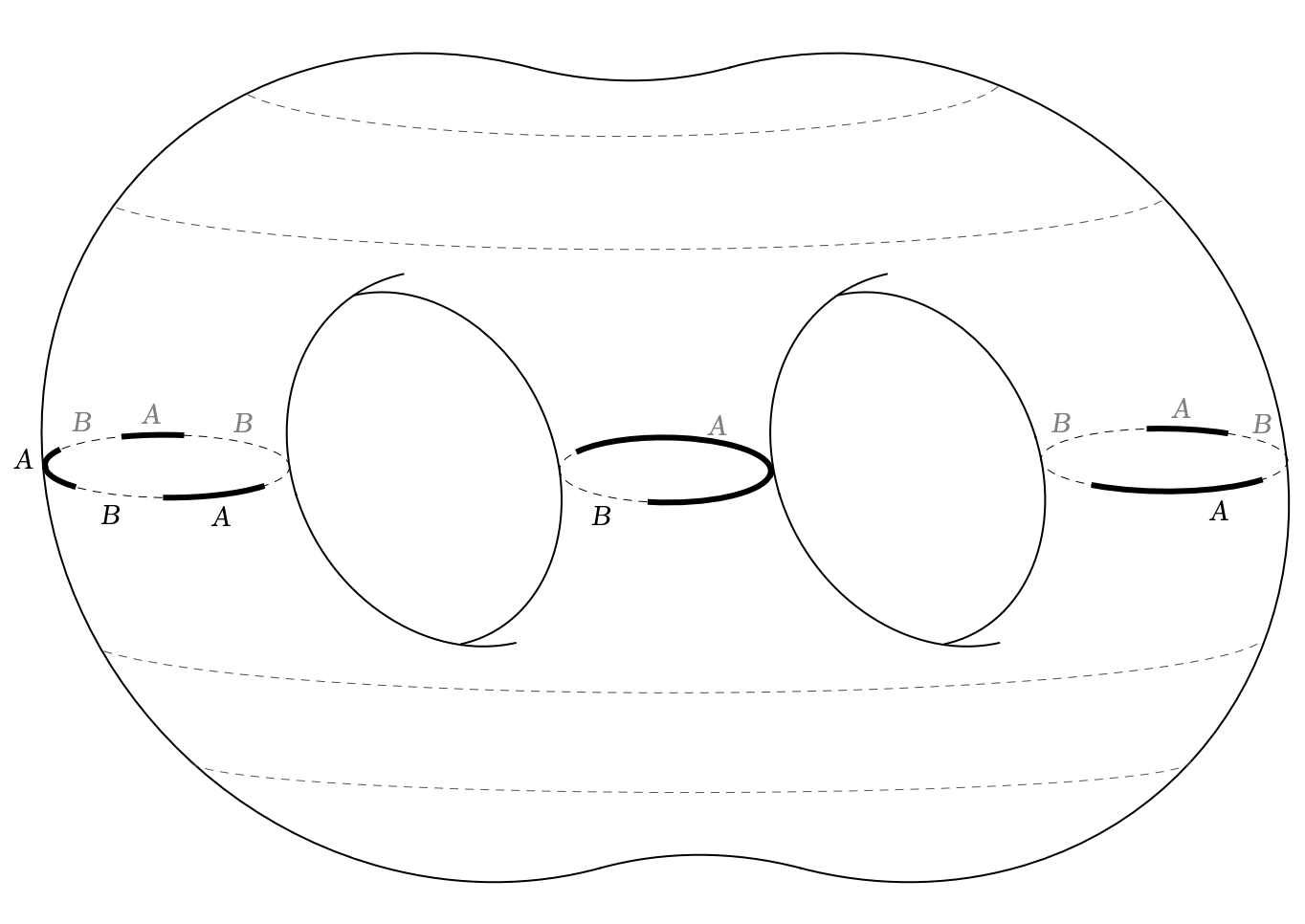}
  \caption{The throat forming in the deconfinement phase after suitable tracing in the imaginary time dimension becomes a closed Riemann surface [27] allowing us to use the Migdal procedure}
  \label{fig:fig3}
\end{figure}
\\
When the surface is being replicated, several tori, with cylindrical throats will be linked together forming a higher genus surface.

\begin{figure}
  \includegraphics[width=160pt]{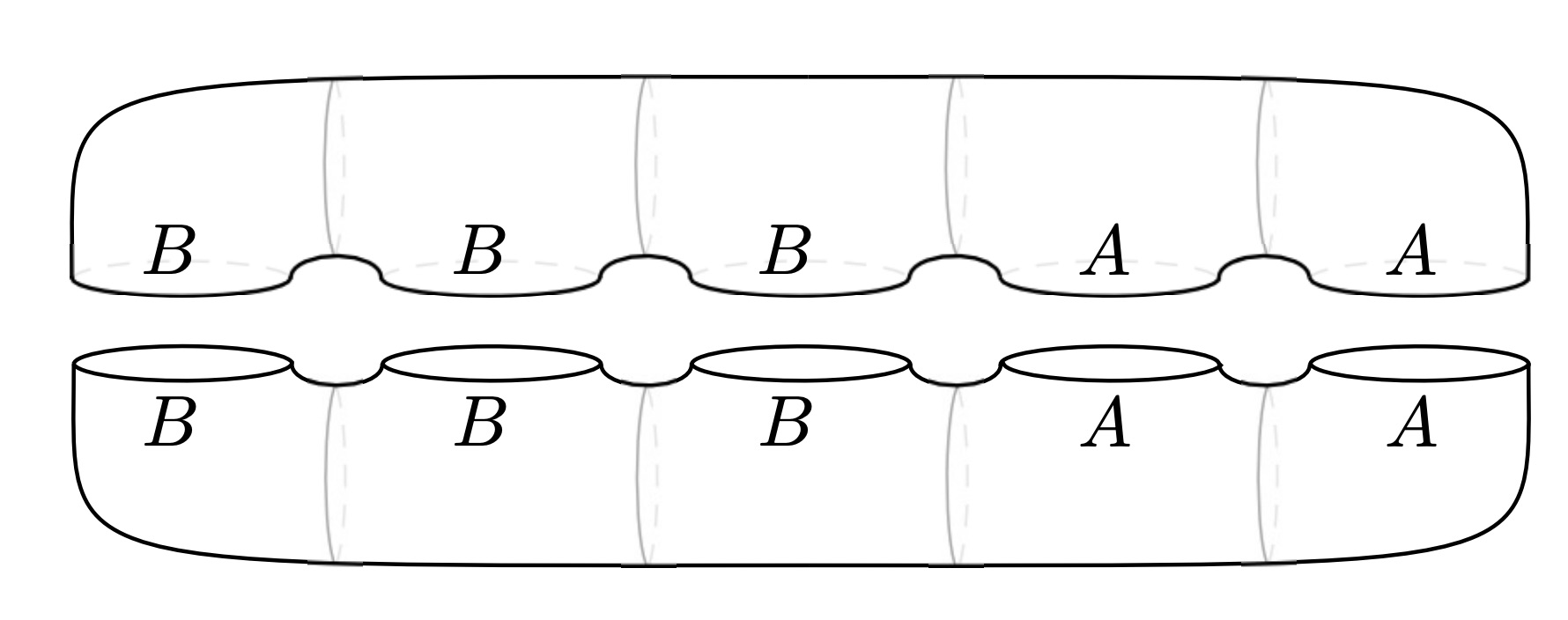}
  \caption{After compactification we obtain a surface of a higher genus which then can be particularised for the desired case. In our case, the genus two would correspond to the situation in which deconfinement exists and there is a non-trivial connection to the region deep inside the bulk, while the genus one will correspond to a confinement phase [27].}
  \label{fig:fig4}
\end{figure}

Our central cylinders forming the connection to the inner regions of the bulk will also be replicated. The formula for the gauge theory partition function depending on the genus of the Riemann surface would therefore be
\begin{equation}
Z=|G|^{g-1}\sum_{R}e^{\rho B_{R}}d_{R}^{2-2g}
\end{equation}
where $|G|$ represents the order of the group while 
\begin{equation}
B_{R}=-\frac{e^{2}c_{2}(R)}{2}
\end{equation}
parametrises a family of quasi-topological quantum field theories. 
With this we can calculate directly via the replica trick the entanglement entropy, considering that in our case the replication also involves additional genera
\begin{widetext}
\begin{equation}
\begin{array}{c}
S_{E}=-lim_{n\rightarrow 1}\frac{d}{dq}\frac{|G|^{[(g-1)q+1]-1}\sum_{R}d_{R}^{2-2[(g-1)q+1]}e^{q \rho B_{R}}}{(|G|^{g-1}\sum_{R'}d_{R'}d_{R'}^{2-2g}e^{\rho B_{R'}})^{q}} =\\

-\sum_{R}\frac{d_{R}^{2-2g}e^{\rho B_{R}}}{\sum_{R'} (d')_{R'}^{2-2g} e^{\rho B_{R'}} } log[ \frac{d_{R}^{2-2g} e^{\rho B_{R}}} {\sum_{R'} d_{R'}^{2-2g} e^{\rho B_{R'}}}]

\end{array}
\end{equation}
\end{widetext}
This formula represents the entanglement entropy obtained in the case of a replica trick calculation following ref. [27] for a two dimensional Yang Mills model. We must be aware that this represents more of a toy model that allows us to compute entanglement entropy in a direct way and probe the effects of higher genus modifications. 
In our case the higher genus is a result of the replication of our baseline configuration which is in the case of the deconfinement phase, after compactifying via the partial trace, a genus two torus. This demands a further limit in our case setting $g\rightarrow 2$. This limit is trivial and using the notation above we obtain 

\begin{widetext}
\begin{equation}
\begin{array}{c}
S_{E}=-\sum_{R}\frac{d_{R}^{-2}e^{\rho B_{R}}}{\sum_{R'} (d')_{R'}^{-2} e^{\rho B_{R'}} } log[ \frac{d_{R}^{-2} e^{\rho B_{R}} } {\sum_{R'} d_{R'}^{-2} e^{\rho B_{R'}}}]
\end{array}
\end{equation}
\end{widetext}
Keeping this in mind, it is interesting to notice that, for the confinement phase description in which the genus-two connectivity is not existent and the D-brane does not connect to the inner regions via the cylinder, we have the $g=1$ case in which the term designated to be the dimension of the representation of the group disappears altogether. This is precisely the case for the confinement phase when using a similar compactification as in the previous case. This will turn the cylinder into a compactified genus-one torus. This would obviously restrict the available area and hence would count a smaller entanglement entropy. 
\par We have therefore computed, following mainly ref. [27], an expression for the entanglement entropy when the connectivity inside the bulk changes and the links of the tensor network can reach deeper inside the bulk. While this calculation involves indeed certain approximations, it does in fact show that a cut in the inner bulk region, limiting the domain where the D-brane can reach, does produce a modification of the available entanglement entropy. There are therefore a series of aspects related to quantum information that are modified when the phase of the system is changed, that are only easily visible when one regards the gravitational dual as well as the associated holographic tensor network. In fact the above formulas link the topology of the regions accessible to the tensor network with accessible entanglement entropy and the possibility of quantum error correction codes of various strengths. The two dimensional Yang Mills model used in the above calculation is of course a simplification that doesn't relate to physical QCD. I do not consider this to be an important issue as the desire of this article is not to create a perfect holographic model of QCD, but instead, to show that in various classes of theories, including, in some limits, also QCD, the access to deeper regions of the bulk depends on the phase of the system, and therefore the quantum error correction capability of a quantum computer will depend on properties that can only be identified from a holographic, gravitationally dual perspective. This toy model shows however, by a different method of analysis, that the same effect of reducing the quantum error correction capability will be obtained if the branes pinch off inside the bulk earlier. 
\par As a task for future research, it is important to evaluate other D-brane systems and their gauge duals for quantum error correction capability, especially in the case in which they can be expanded further inside the bulk and hence have access to more quantum error correction. 
\section{Conclusion}
I gave a quantum information interpretation of the confinement and deconfinement phases in holographic QCD showing that the ability of performing quantum error correction operations in a holographic context is affected by the respective phases. The ability of correcting for boundary erasures is limited in the confinement phase of QCD leading to a new connection between bulk information, its encoding on the boundary, the corresponding erasure errors and various phases of QCD. This study is of course limited by the fact that we do not as for now understand QCD from a holographic perspective, the duality employed being insufficiently studied. However, the new link between quantum information in a holographic context and phase transitions appears to shed new light on an otherwise difficult subject. 

\section{Data Availability Statement}
Data sharing not applicable to this article as no datasets were generated or analysed during the current study.


\begin{thebibliography}{99}
\bibitem{1} J. Zaanen, Y. W. Sun, Y. Liu, K. Schalm, Holographic duality in condensed matter physics, Cambridge University Press, ISBN : 978-1-107-08008-9
\bibitem{2} S. A. Hartnoll, A. Lucas, S. Sachdev, Holographic Quantum Matter, ISBN-13 : 978-0262038430
\bibitem{3} B. Swingle, L. Huijse, S. Sachdev, Phys. Rev. B 90, 045107 (2014)
\bibitem{4} L. Susskind, arxiv hep-th: 1708.03040
\bibitem{5} S. S. Gubser, I. R. Klebanov and A. M. Polyakov, “A semi-classical limit of the gauge/string correspondence,” Nucl. Phys. B636(2002) 99
\bibitem{6} A. May, Tensor networks for dynamic spacetimes, JHEP 2017, 6, pag. 1-27
\bibitem{7} X. Dong, D. Harlow, A. C. Wall, Phys. Rev. Lett. 117(2):021601 (2016)
\bibitem{8} D. J. Gross, R. D. Pisarski, L. G. Yaffe, Rev. Mod. Phys. Vol. 53, 1 (1981)
\bibitem{9} M. Hanada, Y. Matsuo, T. Morita, Nucl. Phys. B 899, pag. 631 (2015)
\bibitem{10} G. Mandal, T. Morita, JHEP 073, 1109 (2011)
\bibitem{11} E. Witten, Adv. Theor. Math. Phys. 2, 505 (1998)
\bibitem{12} M. Kruczenski, D. Mateos, R. C. Myers, D. J. Winters, JHEP, 041, 0405 (2004)
\bibitem{13} R. G. Leigh, Mod. Phys. Lett. A 4, 2767 (1989)
\bibitem{14} J. L. F. Barbon, A. Pasquinucci, Phys. Lett. B 458, 288 (1999)
\bibitem{15} A. Tseytlin, Nucl. Phys. B 501, pag. 41 (1997)
\bibitem{16} D. Gaiotto, N. Itzhaki, L. Rastelli, Nucl. Phys. B Vol. 688, 1, pag, 70 (2004)
\bibitem{17} M. Gutperle, A. Strominger, JHEP 018, 0204 (2002)
\bibitem{18} B. Swingle, Phys. Rev. D 86, 065007 (2012)
\bibitem{19} M. Miyaji, T. Takayanagi, Progr. Theor. Phys. 073B03 (2015)
\bibitem{20} S. Ryu, T. Takayanagi, Phys. Rev. Lett. 96, 181602 (2006)
\bibitem{21} V. E. Hubeny, M. Rangamani, T. Takayanagi, JHEP 062, 2007 (2007)
\bibitem{22} C. H. Bennett, D. P. DiVincenzo, J. A. Smolin, W. K. Wootters, Phys. Rev. A 54 pag. 3824 (1996)
\bibitem{23} C. H. Bennett, G. Brassard, N. D. Mermin, Phys. Rev. Lett. 68, 557 (1992)
\bibitem{24} A. Prudenziati, D. Trancanelli, Phys. Rev. D 96, 026009 (2017)
\bibitem{25} S. Banerjee, Y. Nakaguchi, T. Nishioka, J. High Energ. Phys. 2016, 48 (2016)
\bibitem{26} Zhu-Xi Luo, B. G. Pankovich, Y. Hu, Yong-Shi Wu, Phys. Rev. B 99, 205137 (2019)
\bibitem{27} A. B. Weiss, Studies in: entanglement entropy of two dimensional quasi-topological quantum field theory and geometry of the exact renormalization group and higher spin holography, U. Illinois Rep. 2142/89044 (2015)
\bibitem{28} D. Youm, Mod. Phys. Lett. A 15, pag. 1949-1960 (2000)
\bibitem{29} N. Bao, G. Penington, J. Sorce, A. C. Wall, JHEP vol. 2019, Article number: 69 (2019) 
\bibitem{30} Pastawski, F., Yoshida, B., Harlow, D. et al. Holographic quantum error-correcting codes: toy models for the bulk/boundary correspondence. J. High Energ. Phys. 2015, 149 (2015)
\bibitem{31} G. Araujo-Regado, R. Khan, A. C. Wall, "Cauchy Slice Holography: a new AdS/CFT dictionary", JHEP 2023, 26 (2023)
\bibitem{32} N. Bao, C. Cao, S. M. Carroll, A. Chatwin-Davies, "de Sitter Space as a tensor Network, Cosmic no hair, complementarity and complexity", Phys. Rev. D 96, 123536 (2017)
\bibitem{33} P. Hayden, X-L. Qi, M. Walter, N. Thomas, Z. Yang, S. Nezami, "Holographic duality from random tensor netowkrs", JHEP, 1611, 009 (2016)
\bibitem{34} L. Lepori, L. Dell'Anna, "Long-range topological insulators and weakened bulk-boundary correspondence", New Journal of Physics, Vol. 19, 103030 (2017)
\bibitem{35}M. Alishahiha, S. Banerjee, J. Kames-King, "Complexity via replica trick", JHEP 2022, 181 (2022)
\bibitem{36}R. Vasseur, A. C. Potter, Y-Z. You, A. W. W. Ludwig, "Entanglement transitions from holographic random tensor networks", Phys. Rev. B 100, 134203 (2019)
\bibitem{37}M. Ghodrati, arXiv: 2209.04548v3 "Encoded information of mixed correlations: the views from one dimension higher", (2023)
\bibitem{38}M. Kruczenski, D. Mateos, R. C. Myers, D. J. Winters, "Towards a holographic dual of large $N_{c}$ QCD", JHEP 05 (2004) 041
\bibitem{39} T. Sakai, S. Sugimoto, "Low Energy Hadron Physics in Holographic QCD", Progr. Theor. Phys. Vol 113, Issue 4, pag 843 (2005)
\bibitem{40} S. Sugimoto, K. Takahashi, "QED and string theory", JHEP Vol 2004 JHEP04 (2004)
\bibitem{41} C. Wu, Z. Xiao, D. Zhou, Phys. Rev. D 88, 026016 (2013)

\end{thebibliography}
\end{document}